\documentclass[aps,twocolumn,showpacs,prl,amsmath,amssymb,floatfix,superscriptaddress]{revtex4}

\usepackage{color}
\usepackage{bbm}
\usepackage{graphicx}
\usepackage{dcolumn}
\usepackage{bm}
\usepackage{array}
\usepackage{float}
\usepackage{supertabular}
\usepackage{longtable}
\usepackage{mathrsfs}
\usepackage[T1]{fontenc}
\usepackage{amssymb}
\usepackage[usenames,dvipsnames]{xcolor}
\usepackage{amsmath}
\usepackage{amstext}
\usepackage{latexsym}
\usepackage[colorlinks=true,citecolor=Cerulean,linkcolor=RubineRed,urlcolor=Cerulean]{hyperref}
\usepackage{subfigure}

\begin{document}
\title{Superadiabatic quantum friction suppression in finite-time thermodynamics
}

\author{Shujin Deng}
\affiliation{State Key Laboratory of Precision Spectroscopy, East China Normal University, Shanghai 200062, P. R. China}
\author{Aur\'elia Chenu}
\affiliation{Massachusetts Institute of Technology,
77 Massachusetts Avenue, Cambridge, MA 02139, USA}
\author{Pengpeng Diao}
\affiliation{State Key Laboratory of Precision Spectroscopy, East China Normal University, Shanghai 200062, P. R. China}
\author{Fang Li}
\affiliation{State Key Laboratory of Precision Spectroscopy, East China Normal University, Shanghai 200062, P. R. China}
\author{Shi Yu}
\affiliation{State Key Laboratory of Precision Spectroscopy, East China Normal University, Shanghai 200062, P. R. China}
\author{Ivan Coulamy}
\affiliation{Department of Physics, University of Massachusetts, Boston, MA 02125, USA}
\affiliation{Departamento de F\'isica, Universidade Federal Fluminense, Niter\'oi, RJ, Brazil}
\author{Adolfo del Campo}
\affiliation{Department of Physics, University of Massachusetts, Boston, MA 02125, USA}
\author{Haibin Wu}
\affiliation{State Key Laboratory of Precision Spectroscopy, East China Normal University, Shanghai 200062, P. R. China}
\affiliation{Collaborative Innovation Center of Extreme Optics, Shanxi University, Taiyuan 030006,China}

\newcommand{\be}{\begin{equation}}
\newcommand{\ee}{\end{equation}}
\newcommand{\bea}{\begin{eqnarray}}
\newcommand{\eea}{\end{eqnarray}}

\def\q{{\bf q}}

\def\G{\Gamma}
\def\L{\Lambda}
\def\la{\lambda}
\def\g{\gamma}
\def\al{\alpha}
\def\s{\sigma}
\def\e{\epsilon}
\def\k{\kappa}
\def\ve{\varepsilon}
\def\l{\left}
\def\r{\right}
\def\te{\mbox{e}}
\def\d{{\rm d}}
\def\t{{\rm t}}
\def\K{{\rm K}}
\def\N{{\rm N}}
\def\H{{\rm H}}
\def\la{\langle}
\def\ra{\rangle}
\def\om{\omega}
\def\Om{\Omega}
\def\vep{\varepsilon}
\def\wh{\widehat}
\def\tr{{\rm Tr}}
\def\da{\dagger}
\def\iz{\left}
\def\zi{\right}
\newcommand{\beq}{\begin{equation}}
\newcommand{\eeq}{\end{equation}}
\newcommand{\beqa}{\begin{eqnarray}}
\newcommand{\eeqa}{\end{eqnarray}}
\newcommand{\intf}{\int_{-\infty}^\infty}
\newcommand{\into}{\int_0^\infty}

\begin{abstract}

Optimal performance of thermal machines is reached by suppressing friction.
Friction in quantum thermodynamics results from fast driving schemes that generate nonadiabatic excitations. The far-from-equilibrium dynamics of quantum devices can be tailored by shortcuts to adiabaticity to suppress quantum friction. We experimentally  demonstrate  friction-free  superadiabatic strokes with a trapped unitary Fermi gas as a working substance and establish the equivalence between the superadiabatic mean work and its  adiabatic value.

\end{abstract}

\maketitle


The quest for the optimal  performance of thermal machines  and efficient use of energy resources has motivated the development of finite-time thermodynamics.
At the macroscale, thermodynamic cycles are operated in finite time to enhance the output power, at the expense of inducing friction  and reducing  the efficiency.
Analyzing the tradeoff between efficiency and power has guided efforts in  design and optimization \cite{CA75,FTT84}.
The advent of unprecedented techniques to experimentally control and engineer quantum devices at the nanoscale has shifted the focus to  the quantum domain.
A quantum  engine is an  instance of a thermal machine in which heat \cite{Alicki79,Kosloff84} and other quantum  resources \cite{Scully03,Rossnagel14} can be used to produce work.
The experimental realization of a single-atom heat engine  \cite{Rossnagel16} and a quantum absorption refrigerator  \cite{Maslennikov17} have been demonstrated using trapped ions.

\begin{figure*}
  \begin{minipage}{.2\textwidth}
    \includegraphics[angle=0,width=\textwidth,height=4 in]
    {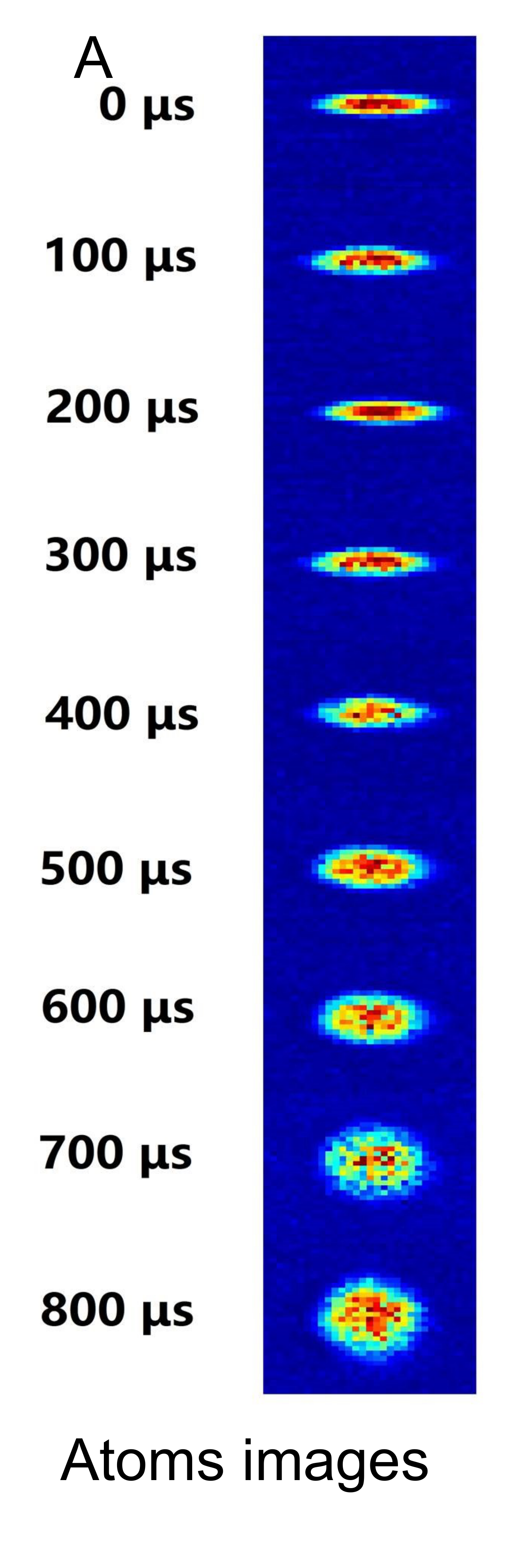}
  \end{minipage}
  \begin{minipage}{.35\textwidth}
    \includegraphics[angle=0,width=\textwidth,height=4 in]
    {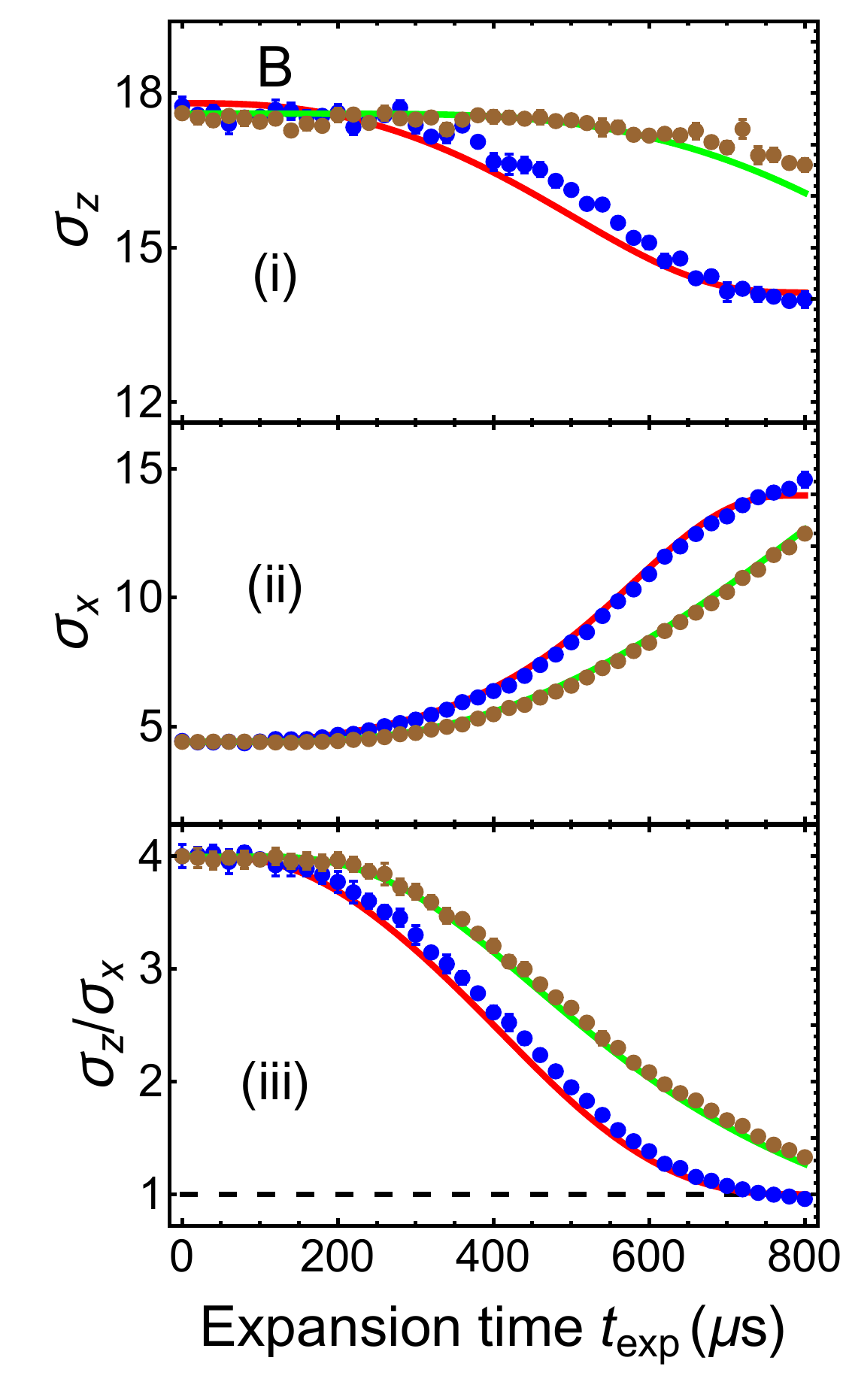}
  \end{minipage}
  \begin{minipage}{.35\textwidth}
    \includegraphics[angle=0,width=\textwidth,height=4 in]
    {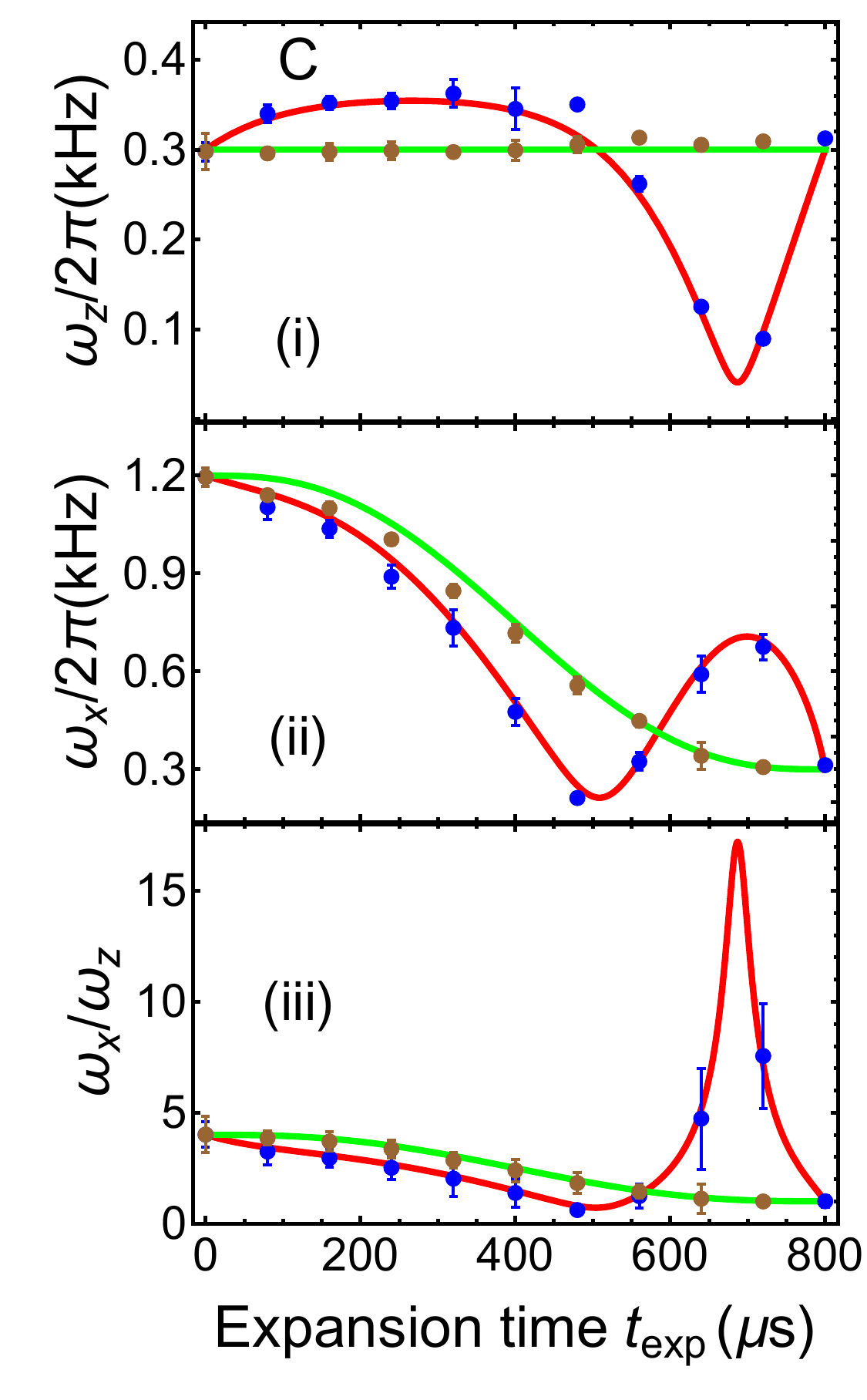}
  \end{minipage}
    \caption{The expansion stroke of the unitary Fermi gas for STA of LCD and reference driving case with $b_{f,x}=4$, keeping $b_{f,y}=b_{f,z}=1$ in $t_f=800 \mu s$:  the atoms images of LCD driving (A), the measured cloud size and cloud aspect ratio (B), and  the measured frequencies and frequency aspect ratio (C).  Blue dots and brown dots are the measured results for LCD and reference driving case, respectively. The red line and green line are theoretical predictions. The black dashed line denotes for the aspect ratio of 1, indicating an isotropic trap for Fermi gas. Error bars represent the standard deviation of the statistic.}\label{fig1}
\end{figure*}

In the quantum domain, the adiabatic theorem \cite{Kato50} dictates that excitations are formed during  fast driving of the working substance,  leading to the  emergence of  quantum friction.
Trading efficiency and power  remains a predominant strategy in finite-time quantum thermodynamics, e.g., of ground-state cooling \cite{GK92,YK06,Shiraishi16}.
In parallel, new efforts have been devoted to completely suppress friction in finite-time quantum processes.
A systematic way of achieving this goal is provided by shortcuts to adiabaticity (STA): fast nonadiabatic processes that reproduce  adiabatic dynamics, e.g., in the preparation of a target state \cite{STAR}. The use of STA provides a disruptive approach in finite-time thermodynamics and has motivated proposals for superadiabatic thermal machines, operating at maximum efficiency and arbitrarily high output power \cite{Deng13,delcampo14,Beau16,Abah16}. STA engineering is facilitated by counterdiabatic driving \cite{demirplak03,Berry09}, whereby an auxiliary control field speeds up the evolution of the system through  an adiabatic reference trajectory in a prescheduled amount of time. Experimental demonstrations of counter-diabatic driving have focused on effectively single-particle systems at zero temperature \cite{expCD1,expCD2,expCD3,expCD4}.  Tailoring excitation dynamics is expected to be a daunting task in complex systems. Yet efficient quantum thermal machines offering scalability require the superadiabatic control of the finite-time thermodynamics in many-particle systems \cite{Beau16}.
We report the suppression of quantum friction in the finite-time thermodynamics of a strongly-coupled  quantum fluid. In our experiment, we implement friction-free
superadiabatic strokes with a unitary Fermi gas in an anisotropic time-dependent trap as a working medium.

The unitary regime is reached when the scattering length governing the short-range interactions in a spin-$1/2$ ultracold Fermi gas at resonance greatly surpasses the interparticle spacing. This strongly-interacting state of matter  is described by a  nonrelativistic conformal field theory   \cite{DTSon}. The controllability of the external trap potential and interatomic interactions in this system allows for the preparation of well-defined many-body states and the precise engineering of time-dependent Hamiltonians. This provides unprecedented opportunities for studying strongly interacting nonequilibrium phenomena \cite{BCSBEC}. 
In particular, the emergent scale-invariance symmetry at unitarity facilitates the realization of superadiabatic strokes by the counterdiabatic driving scheme.
The work done in a stroke induced by a modulation of the trap frequency  can be described by a probability distribution \cite{Talkner07}. The mean work equals the difference between the energy of the cloud brought out of equilibrium at the end of the stroke and its initial  equilibrium value. Therefore,  work done by the cloud is negative while work done on it is positive. For a unitary dynamics, the emergent scaling symmetry at  strong coupling \cite{anisotropicexpansion}  dictates the evolution of the nonadiabatic energy in terms of the cloud density profile. The density profile can be characterized using the collective coordinate operators
\beqa
\hat{R}_x^2=\sum_{i=1}^N \hat{x}_i^2,\quad \hat{R}_y^2=\sum_{i=1}^N \hat{y}_i^2, \quad \hat{R}_z^2=\sum_{i=1}^N \hat{z}_i^2.
\eeqa
 Their expectation values  determine the shape of the atomic cloud via  the scaling factors $b^2_j(t)=\la R_j^2(t)\ra/\la R_j^2(0)\ra$  in each axis $j=x,y,z$. Their evolution in time is governed by the coupled equations \cite{3DFermi}
\begin{equation}
\ddot{b}_j(t)+\omega_j^2(t)b_j=\frac{\omega_{j,0}^2}{b_j\Gamma^{2/3}},\quad j=x,y,z\, , \label{eq1}
\end{equation}
 where $\Gamma=b_x(t)b_y(t)b_z(t)$ is the scaling volume. Under slow driving, the adiabatic scaling factor reads
\begin{align}
b_{j,ad}(t)&=\frac{1}{\Gamma_{ad}^{1/3}}\frac{\omega_{j,0}}{\omega_j(t)},\label{eq2}\\
\Gamma_{ad}(t)&=\prod_jb_{j,ad}(t).\label{eq3}
\end{align}
In characterizing the finite-time thermodynamics of isolated quantum systems, the ratio of the nonadiabatic and adiabatic mean energies plays a crucial role and is known as the nonadiabatic factor
$Q^*(t)=\la\hat{H}(t)\ra/\la\hat{H}(t)\ra_{ad}$ \cite{Husimi53,Jaramillo16}.
For a unitary anisotropic Fermi gas,   the nonadiabatic factor is given by (see supplementary text)
\beqa \label{eq:Qstar}
Q^*(t)=\Gamma_{ad}(t)^{2/3}\left[\frac{1}{2\Gamma^{2/3}}+\frac{1}{6}\sum_{j=x,y,z}\frac{\dot{b}_j^2+\om_j^2(t)b_j^2}{\om_{j,0}^2}\right],
\eeqa
and determines the nonadiabatic mean work
\beqa
\la W(t)\ra=(Q^*(t)/\Gamma_{ad}^{2/3}(t)-1)\la H(0)\ra.
\eeqa

Quantum friction is evidenced during dynamical processes with values of  $Q^*(t)>1$.
To suppress friction, the counterdiabatic driving technique \cite{demirplak03,Berry09} can be exploited in combination with dynamical scaling laws  \cite{delcampo13,DJD14}  to set $Q^*=1$.
We refer to such superadiabatic control as local counterdiabatic driving (LCD).  To implement it, we first identify a desirable reference trajectory of the trap frequencies $\omega_j(t)$.  We next design a STA protocol with modified frequencies $\Omega_j(t)$  that reproduces in an arbitrary prescheduled time $t_f$ the  final state that would correspond to the adiabatic evolution for $\omega_j(t)$.
To this end, we choose a reference  modulation of the trap, specifically
\begin{align}
\omega_j(t)=\omega_j(0)\{1+(b_{f,j}-1)[10\tau^3-15\tau^4+6\tau^5]\},\label{eq7}
\end{align}
where $\tau=t/t_f$ and $b_{f,j} = \omega_{j}(t_f)/\omega_j(0)$ is set by the ratio of the initial and final target trap frequency. The engineering of a STA by LCD  requires the nonadiabatic  modulation of the trap frequencies (see supplementary text)
\begin{align}
\Omega_j^2(t)&=\omega_{j}^2-2\left(\frac{\dot{\omega}_j}{\omega_j}\right)^2+\frac{\ddot{\omega}_j}{\omega_j}
+\frac{1}{4}\left(\frac{\dot{\nu}}{\nu}\right)^2-\frac{1}{2}\frac{\ddot{\nu}}{\nu}+\frac{\dot{\omega}_j\dot{\nu}}{\omega_j\nu},\label{eq6}
\end{align}
where $\nu(t)=[\omega_x(t)\omega_y(t)\omega_z(t)]^{1/3}$ denotes the  geometric mean frequency. The frequencies $\Omega_j(t)$ thus satisfy the desired boundary conditions, $\Omega_j(0)=\omega_j(0)$, $\Omega_j(t_f)=\omega_j(t_f)$,  and $\dot{\omega}_j(0)=\ddot{\omega}_j(t_f)=0$, while ensuring $Q^*(t_f)=1$. While the LCD dynamics is nonadiabatic at intermediate stages, quantum friction is thus suppressed upon completion of the stroke of arbitrary duration $t_f$.

Our experiment probes the nonadiabatic expansion dynamics in an anisotropically-trapped unitary quantum gas, a balanced mixture of $^6$Li fermions in the lowest two hyperfine states  $|\!\!\uparrow\rangle\equiv|F=1/2, M_F=-1/2\rangle$ and $|\!\!\downarrow\rangle\equiv|F=1/2, M_F= 1/2\rangle$. The experimental setup is similar to that in  Ref.~\cite{Wu1,Wu2}, with a new configuration of the dipole trap consisting in  an elliptic beam generated by a cylindrical lens along the $x$-axis and a nearly-ideal Gaussian beam along the $z$-axis (see  supplementary text). The resulting potential has a cylindrical symmetry around $x$. This trap facilitates the control of the anisotropy and  geometric frequency. Fermionic atoms are loaded into a cross-dipole trap used for evaporative cooling. A Feshbach resonance is used to tune the interactomic interaction  to the unitary limit, reached at $B=832\,$G. The system is initially prepared in a stationary state of a normal fluid, with $\omega_x(0)=2\pi\times1200$ Hz and $\omega_y(0)=\omega_z(0)=2\pi\times300$ Hz. The initial energy of Fermi gas at unitarity is $E=0.8 (0.1)\,E_F$, corresponding to a temperature $T=0.25(0.02)\,T_F$, where $E_F$ and $T_F$ are the Fermi energy and temperature of an ideal Fermi gas, respectively.

The time-dependent trap frequencies and trap anisotropy are precisely controlled according to equation (\ref{eq7}) for the reference driving, and equation (\ref{eq6}) for the LCD driving.   To monitor the evolution along the process, after a chosen expansion time  with the trap turned on, the trap is completely turned off and the cloud is probed by standard resonant absorption imaging techniques after a time-of-flight expansion time $t_{\rm tof}=600\,\mu s$. Each data point is obtained from averaging $5$ runs of measurements with identical parameters. The time-of-flight density profile along the axial (radial) direction is fitted by a Gaussian function, from which we obtain the observed cloud size $\sigma_{z,\rm obs}$($\sigma_{r,{\rm obs}}$). The \textit{in-situ} cloud size  $\sigma_z$ ($\sigma_r$) and scaling  factors $b_j$ during the STA are obtained from the observed value $\sigma_{z,{\rm obs}}$ ($\sigma_{r,{\rm obs}}$) scaled by factor evaluated from the  hydrodynamic  expansion equation during a time $t_{\rm tof}$ (see supplementary text).

Figure \ref{fig1} shows the expansion stroke of the unitary Fermi gas with $b_{f,x}=1/4$, $b_{f,y}=b_{f,z}=1$ in $t_f=800 \mu s$. The  trap frequencies are changed from the initial values $\omega_x(0)=2\pi\times1200$ Hz and $\omega_y(0)=\omega_z(0)=2\pi\times300$ Hz to the target values $\omega_x(t_f)=2\pi\times300$ Hz and $\omega_y(t_f)=\omega_z(t_f)=2\pi\times300$ Hz in 800 $\mu$s.  The Fermi gas is initially confined in an anisotropic harmonic trap with a frequency aspect ratio of 4. Due to the engineering of frequencies, the size of the cloud gas is mostly expanded in the $x$ axis,  the changes in the $y$ and $z$ directions being small (Fig. \ref{fig1}B) during the driving processes.
While the Fermi gas is anisotropic during the nonadiabatic, LCD drive, it becomes isotropic at  the final target state, in which both the frequency and cloud size aspect ratio are close to 1; see Fig. 1B(iii) and Fig. 1C(iii).
\begin{figure}[t]
\centering
\subfigure{ \label{fig2A}
\includegraphics[width=8cm]{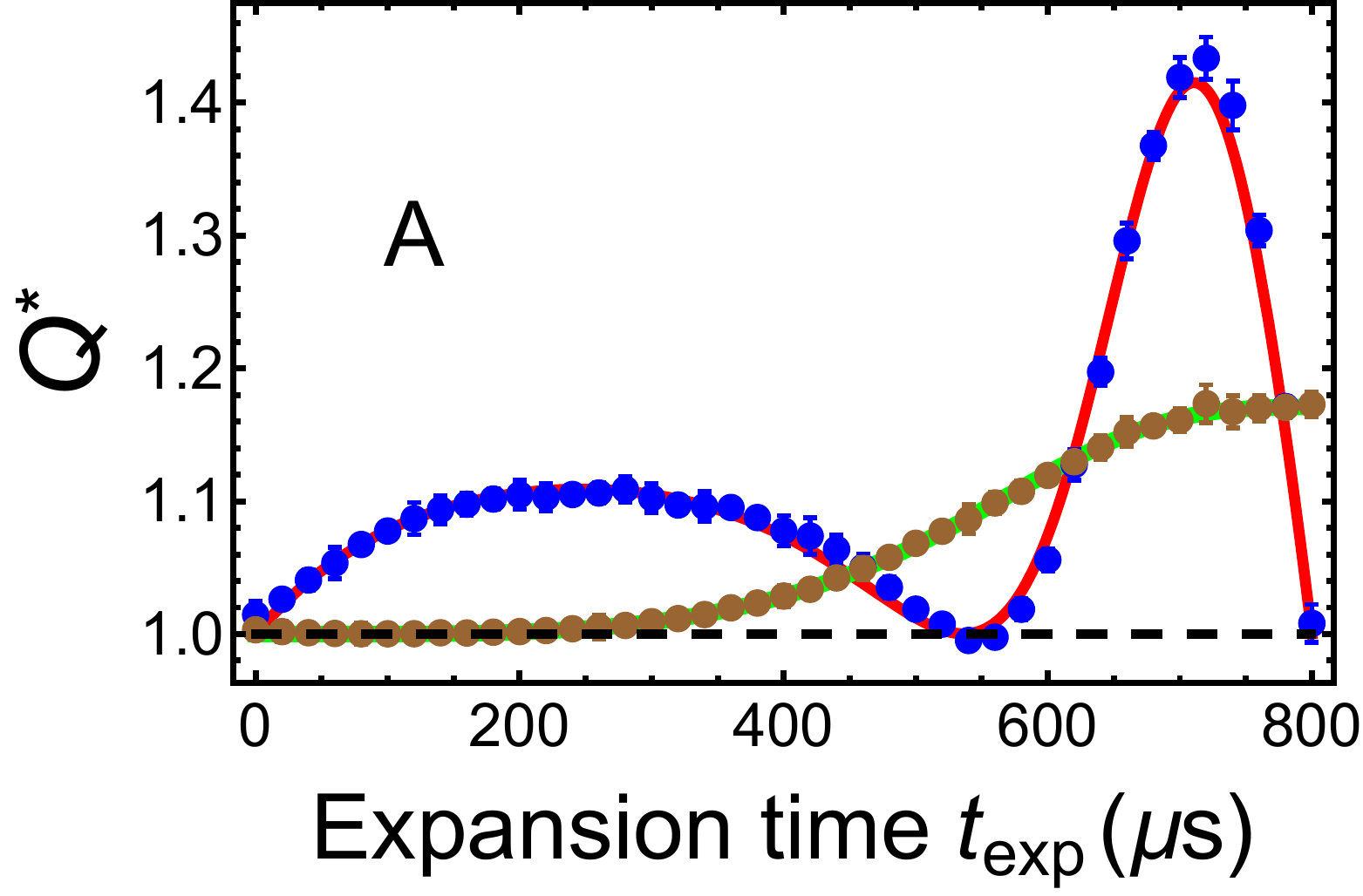}
}
\hspace{2.0cm}
\subfigure{ \label{fig2B}
\includegraphics[width=8cm]{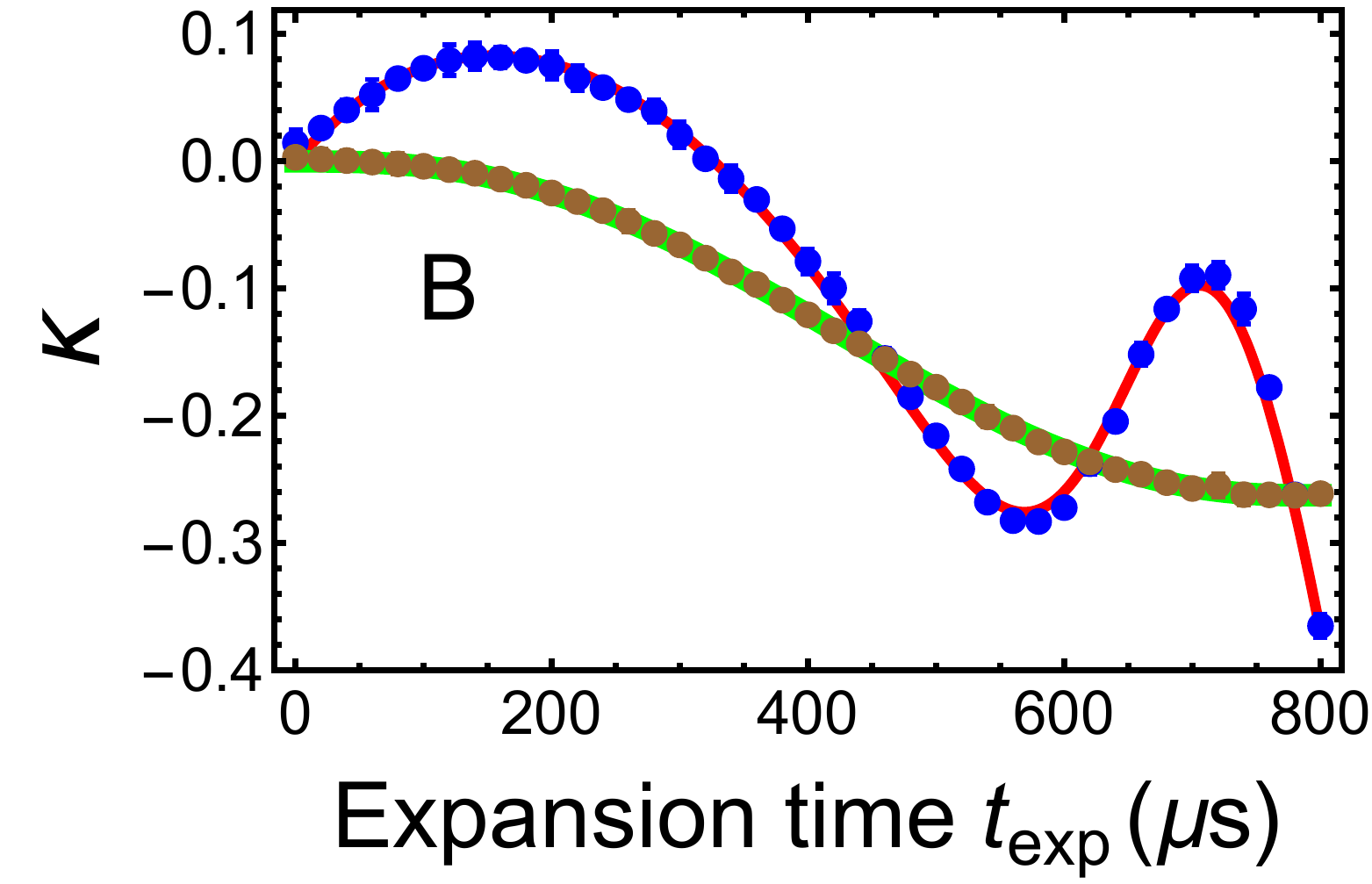}
}
\caption{Nonadiabatic factor $Q^*$ (A) and the mean work (B) in an expansion stroke. Blue and brown dots denote the measured data for STA of local counter-diabatic driving and nonadiabatic reference driving case while  the red  and green lines are the corresponding theoretical predictions. $\kappa(t)=\la W(t)\ra/\langle\hat{H}(0)\rangle$ is the ratio of the mean work and the initial energy. The black dashed line represents $Q*=1$ where there is no quantum friction. Error bars represent the standard deviation of the statistic.}\label{fig2}
\end{figure}

\begin{figure*}
  \begin{minipage}{.16\textwidth}
    \includegraphics[angle=0,width=\textwidth,height=1.8 in]
    {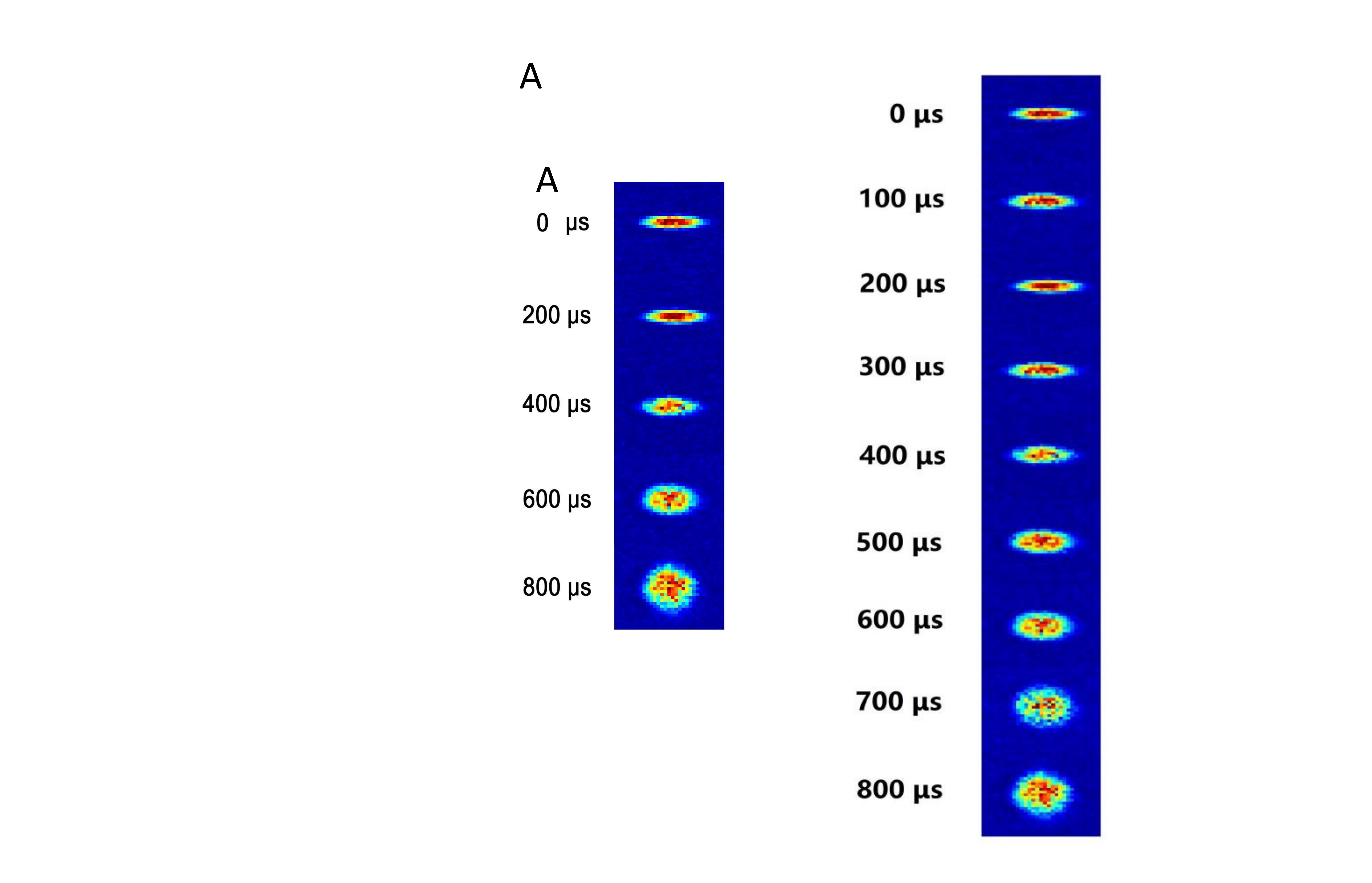}
  \end{minipage}
  \begin{minipage}{.35\textwidth}
    \includegraphics[angle=0,width=\textwidth,height=1.8 in]
    {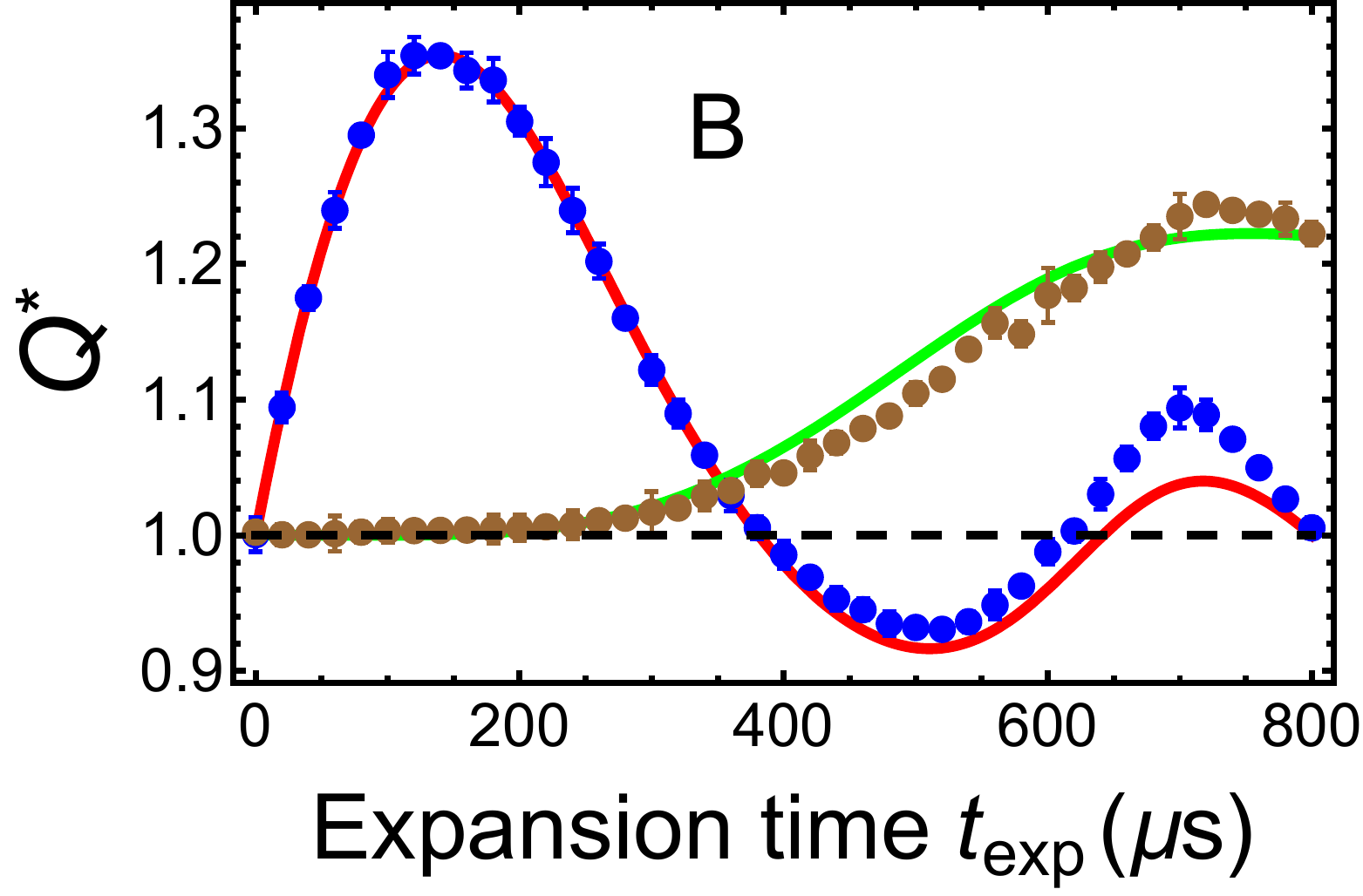}
  \end{minipage}
  \begin{minipage}{.35\textwidth}
    \includegraphics[angle=0,width=\textwidth,height=1.8 in]
    {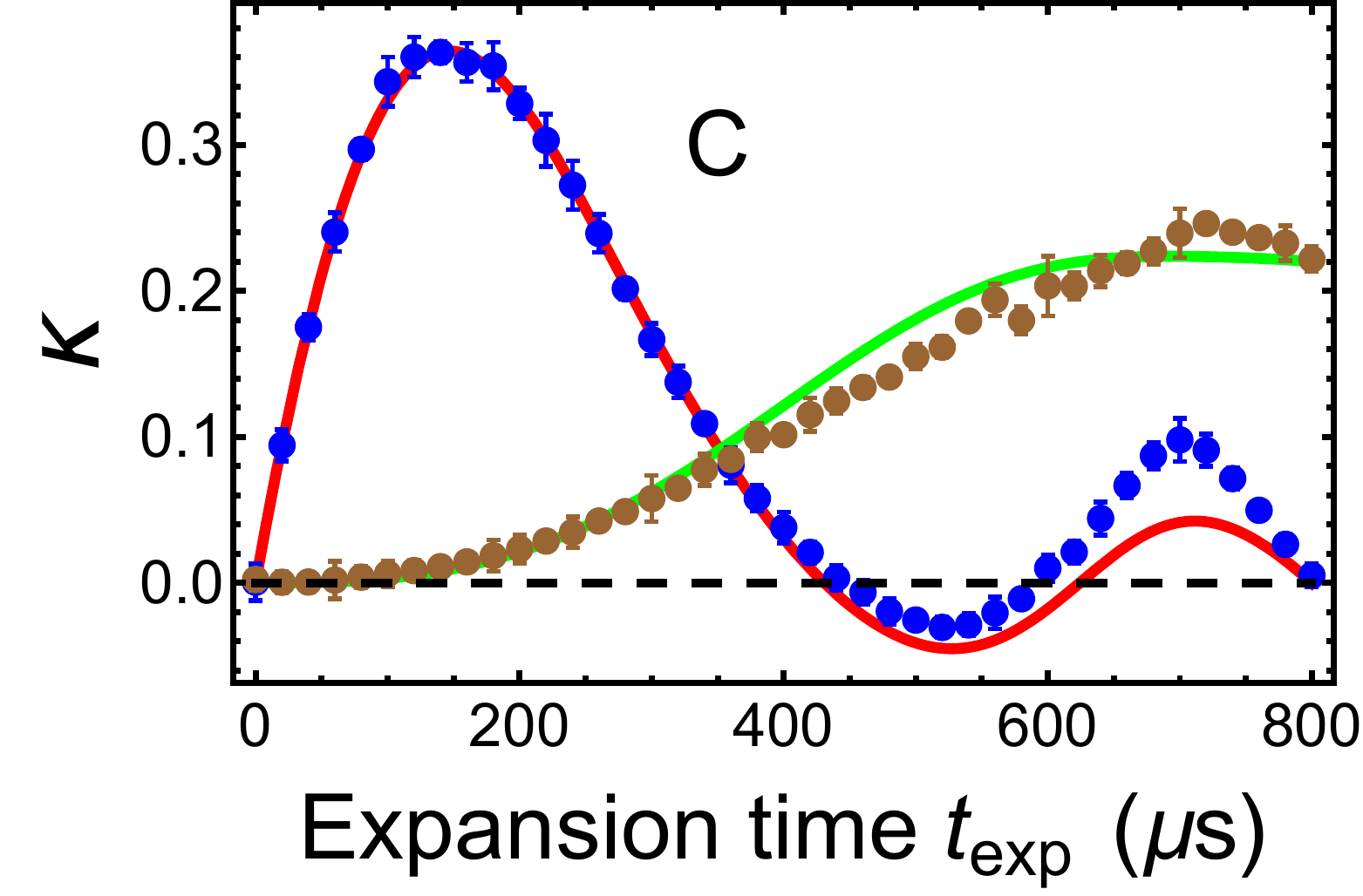}
  \end{minipage}
    \caption{\label{fig3}
    Cloud expansion images (A),  nonadiabatic factor $Q^*$ (B) and  mean work (C) of unitary Fermi gas during a change of the cloud aspect ratio that preserves the geometric mean frequency, $\nu(t_f)=\nu(0)$. Blue  and brown dots represent the measured results for STA of LCD and nonadiabatic driving case, respectively. Red and green lines denote the corresponding theoretical predictions. Black dashed line denotes for the friction-free $Q^*(t_f)=1$ and brown dashed line represents for the zero mean work $\kappa=0$. Error bars represent the standard deviation of the statistic.
}
\end{figure*}

The measured nonadiabatic factor $Q^*(t)$ and mean work during the expansion stroke are shown in Fig. \ref{fig2}. For the reference  driving,  the $Q^*(t)$ of the strongly interacting Fermi gas  monotonically increases, witnessing quantum friction induced by  excitations that emerge from the nonadiabatic dynamics. At time $t_f=800\mu s$, $Q^*(t_f)$ approaches to 1.17 for the experimental parameters  (brown dots in Fig. \ref{fig2}A). Quantum friction  decreases the mean work of the stroke, $\kappa(t)=\la W(t)\ra/\langle\hat{H}(0)\rangle$ (see Fig. \ref{fig2}B).
By contrast,  quantum friction is greatly suppressed with  the local counterdiabatic driving scheme. Indeed, $Q^*(t)$  reduces to 1 after completion of the superadiabatic stroke (blue dots in Fig. \ref{fig2}A), revealing that it is a friction-free  process.  Using  STA by LCD thus allow reaching $Q^*(t_f)=1$, enhancing  the  work output. Its maximum value  is reached  at $t_f$ and is determined by the geometric mean frequency according to
\begin{equation}
\la W(t_f)\ra_{\rm LCD}=\bigg(\bigg(\frac{\nu(t_f)}{\nu(0)}\bigg)^{1/3}-1\bigg)\langle\hat{H}(0)\rangle.\label{limit}
\end{equation}
 The measured value  of the mean work $\kappa$ in the experiment is about  $-0.37$  in units of $\langle\hat{H}(0)\rangle$ (blue dots in Fig. \ref{fig2}B), which is in very good agreement with the theoretical prediction, $(2^{-2/3}-1)=-0.370$. Compared to the case of reference driving process, the mean work in the the local counterdiabatic driving process is increased by nearly 42.3$\%$.

As seen from Eq.~\ref{limit}, in a STA by LCD  quantum friction is suppressed and the final mean work  depends only on  the ratio of geometric mean frequency between the initial state and the target state.
For processes satisfying $\nu(t_f)=\nu(0)$, such  as a change of the anisotropy of the cloud, the mean work vanishes. We verify this prediction experimentally. The strongly interacting Fermi gas is first prepared in a quantum state with frequencies of $\omega_x(0)=2\pi\times1200$ Hz and $\omega_y(0)=\omega_z(0)=2\pi\times300$ Hz. Then, following by Eq. \ref{eq7}, the trap frequencies are changed to  $\omega_x(t_f)=\omega_y(t_f)=\omega_z(t_f)=2\pi\times(\prod_j\omega_j(0))^{1/3}=2\pi\times476.22$ Hz, modifying the aspect ratio of the cloud. The measured nonadiabatic factor $Q^*(t)$, mean work  and  cloud expansion images are presented in Fig. \ref{fig3}. It is apparent that there is no friction $Q^*(t_f)=1$ (blue dots in Fig. \ref{fig3}A) and that the mean work vanishes $W(t_f)=0$ (blue dots in Fig. \ref{fig3}B) when manipulating the quantum state using STA by LCD. For the nonadiabatic reference driving,  quantum friction is however induced and causes  $Q^*(t_f)>1$ (brown dots in Fig. \ref{fig3}A), signaling  energy dispersion in the final state. The mean work is then increased to a positive value of 1.22$\langle\hat{H}(0)\rangle$ (brown dots in Fig. \ref{fig3}B).

A global measure of the nonadiabatic character of a superadiabatic stroke can be quantified by the time-average deviation of the mean work  $\la W(t)\ra$ from the adiabatic value $\la W_{ad}(t)\ra$ as a function of the total time $t_f$
\beqa
\delta W=\frac{1}{t_f}\int_0^{t_f}dt\left[\la W(t)\ra-\la W_{ad}(t)\ra\right].
\eeqa
 In the experiment, we measure the mean work deviation in both a STA based on LCD driving and a nonadiabatic reference protocol for the expansion  strokes.  The initial and final geometric frequencies are fixed to $4.8$ and $1.2$, respectively.  The trap frequencies are changed from the initial frequencies $\omega_x=2 \pi \times 1440$ Hz, $\omega_y=\omega_z=2 \pi \times 300$ Hz to  $\omega_x=2 \pi \times 360$ Hz, $\omega_y=\omega_z=2 \pi \times 300$ Hz in different time $t_f$. For STA by LCD driving, the mean work $\la W(t_f)\ra$ does not change  no matter how short the time $t_f$ is. However,  $\la W(t)\ra$ at $t<t_f$ keeps changing to reflect how large the excess work needed to implement STA at different stages. The measured results are shown in Fig. \ref{fig4}.  The shorter the time $t_f$, the greater the deviation of mean work $\delta W$. The solid curves are the fits with a power-law $1/t_f^2$ as a function of the duration of the stroke $t_f$.

\begin{figure}[t]
\centering
\includegraphics[width=8cm]{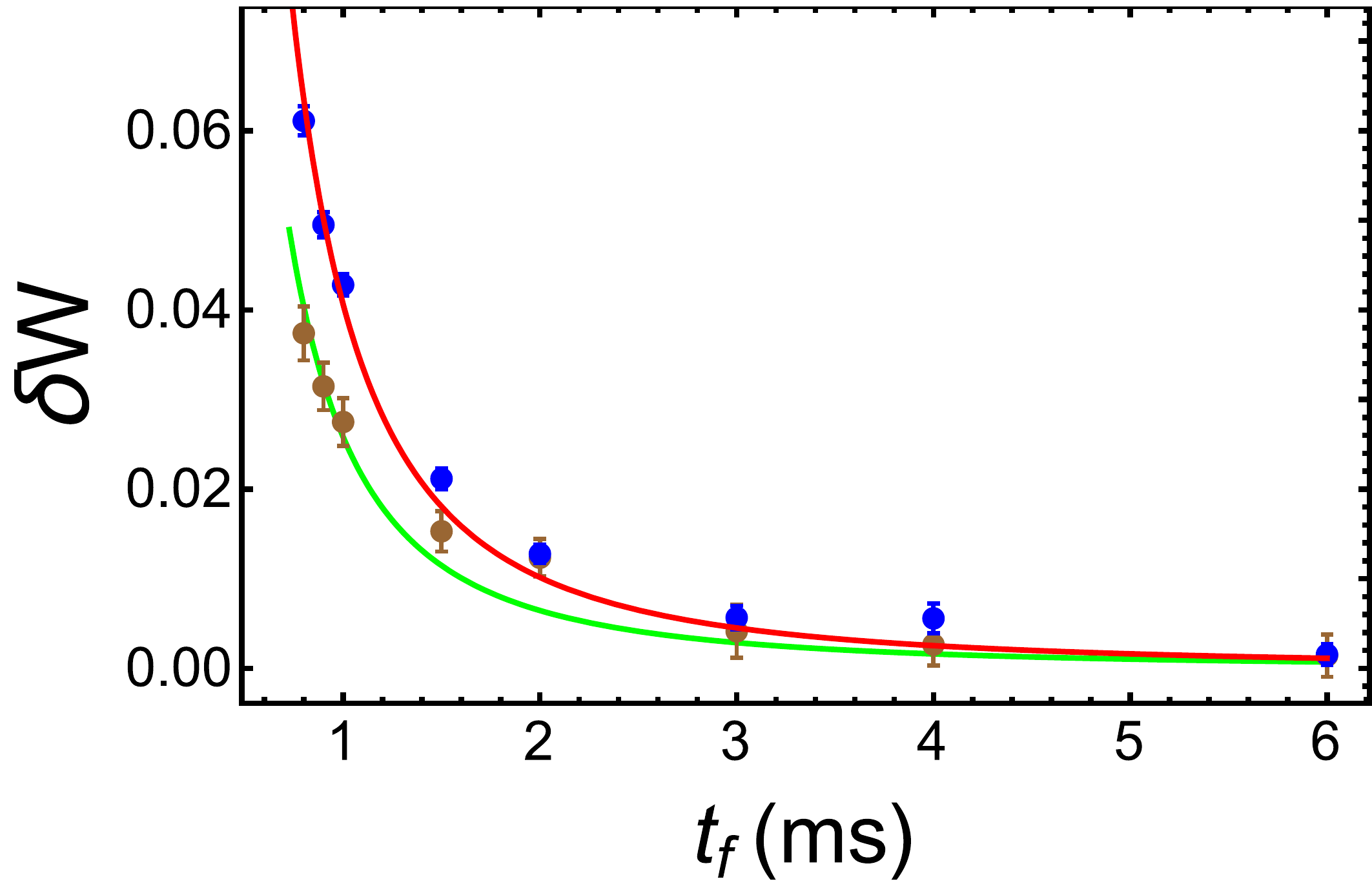}
\caption{Time-averaged mean work deviation $\delta W$  in units of $\langle H(0)\rangle$ as a function of $t_f$. Blue and brown dots denote the measured data for STA  by LCD driving and the nonadiabatic reference driving case,  while  the red  and green lines are the fits with a function of $1/t_f^2$. Error bars represent the standard deviation of the measurement data.}\label{fig4}
\end{figure}

The experiments presented here demonstrate the suppression of quantum friction and enhancement of the mean work output in superadiabatic strokes with a strongly interacting quantum fluid  as a working medium. The control of the finite-time quantum thermodynamics is achieved using shortcuts to adiabaticity that exploit the emergent scale-invariance of a unitary Fermi gas. In combination with cooling and heating steps, superadiabatic strokes can be used to engineer quantum heat engines \cite{Deng13,delcampo14,Beau16,Salamon09,Abah16} and refrigerators \cite{Rezek09,Stefanatos10,Hoffmann11}, e.g., based on a quantum Otto cycle,  that operate at maximum efficiency with high output and cooling power. 

{\it Acknowlegments.---}
This research is supported by the National Key Research and Development Program of China (grant no.2017YFA0304201), National Natural Science Foundation of China (NSFC) (grant nos. 11374101 and 91536112), Program of Shanghai Subject Chief Scientist(17XD1401500), UMass Boston (project P20150000029279)  and the John Templeton Foundation.

\newpage
\widetext
\section{Supplemental Material}
\label{sec:supplement}

\section{Finite-time thermodynamics of a unitary Fermi gas in a time-dependent anisotropic trap}
The Hamiltonian of a unitary Fermi gas confined in a 3D anisotropic harmonic trap whose frequency is time dependent is given by
\begin{equation}
\label{Hsystem}
\hat{H}(t)=\sum_{i=1}^N\left[-\frac{\hbar^2}{2m}\nabla_i^2+\frac{1}{2}m\left(\omega_x^2(t)x_i^2+\omega_y^2(t)y_i^2+\omega_z^2(t)z_i^2\right)\right]+\sum_{i<j}U({\bf r}_i-{\bf r}_j),
\end{equation}
where $\omega_j$ ($j=x,y,z$) is the corresponding trapping frequency along the $j$-axis. The short-range pair-wise interaction potential becomes a homogeneous function of degree $-2$ in the unitary limit, satisfying
\begin{equation}
U(\lambda{\bf r})=\frac{1}{\lambda^2}U({\bf r}).
\end{equation}
In what follows it is convenient to rewrite the total hamiltonian $\hat{H}(t)$ as the sum
\begin{equation}
\hat{H}(t)=\hat{H}_0+\hat{H}_{\rm trap}(t),
\end{equation}
where
\begin{eqnarray}
\hat{H}_0&=&\sum_{i=1}^{N}\left(-\frac{\hbar^2}{2m}\nabla_i^2\right)+\sum_{i<j}U(\mathbf{r}_i-\mathbf{r}_j),\\
\hat{H}_{\rm trap}(t)&=&\frac{1}{2}m\sum_{i=1}^{N}\left[\omega_x^2(t)x_i^2+\omega_y^2(t)y_i^2+\omega_z^2(t)z_i^2\right].\label{eq1}
\end{eqnarray}
For an exact treatment of the finite-time thermodynamics we compute the nonadiabatic evolution of the mean work.  To this end, we note that
\beqa
\frac{d}{dt}\hat{H}_0&=&\frac{1}{i\hbar}\left[\hat{H}_0,\hat{H}(t)\right]\\
&=&\frac{1}{i\hbar}\left[\hat{H}(t)-\hat{H}_{\rm trap}(t),\hat{H}(t)\right]\\
&=&-\frac{1}{2}m\sum_{i=1}^N\left(\om_x^2(t)\frac{d\hat{x}_i^2}{dt}+\om_y^2(t)\frac{d\hat{y}_i^2}{dt}+\om_z^2(t)\frac{d\hat{z}_i^2}{dt}\right). \label{eq:dHdt}
\eeqa
We introduce the collective coordinate operators
\begin{equation}
\hat{R}_x^2=\sum_{i=1}^N \hat{x}_i^2,\quad \hat{R}_y^2=\sum_{i=1}^N \hat{y}_i^2, \quad \hat{R}_z^2=\sum_{i=1}^N \hat{z}_i^2,
\end{equation}
and define via their expectation values  the scaling factors 
\beqa
b_j(t)=\left[\frac{\langle \hat{R}_j^2(t)\ra}{\la \hat{R}_j^2(0)\ra}\right]^{1/2}, \quad j=x,y,z.
\eeqa

For an equilibrium state, we use Hellman-Feyman theorem and the scaling property of the Hamiltonian to find
\begin{equation}
\la \hat{R}_j^2(0)\ra=\frac{1}{3 m\om_{j,0}^2}\la \hat{H}(0)\ra, \quad j=x,y,z.
\end{equation}

The expectation value of Eq. (\ref{eq:dHdt}) can thus be written as
\beqa
\frac{d}{dt}\la \hat{H}_0\ra
&=&-\frac{1}{2}m\sum_{j=x,y,z}\om_j^2(t)\frac{d\la R_j^2(t)\ra}{dt}\\
&=&-\frac{1}{6}\sum_{j=x,y,z}\frac{\om_j^2(t)}{\om_{j,0}^2}\frac{db_j^2}{dt}\la\hat{H}(0)\ra\\
&=&-\frac{1}{3}\sum_{j=x,y,z}\frac{\om_j^2(t)}{\om_{j,0}^2}b_j\dot{b}_j\la\hat{H}(0)\ra.
\label{meanen0}
\eeqa

For a 3D unitary Fermi gas, the scaling factors are coupled  via the evolution equations
\beqa
\label{eceq}
\ddot{b}_j+\om_j^2(t)b_j=\frac{\om_{j,0}^2}{b_j\Gamma(t)^{2/3}}, \quad j=x,y,z,
\eeqa
where the volume scaling factor is given by $\Gamma(t)=\prod_jb_j(t)$.

Combining (\ref{eceq}) and (\ref{meanen0}), we rewrite the rate of change of the expectation value of $\hat{H}_0$ as
\beqa
\frac{d}{dt}\la \hat{H}_0\ra&=&\frac{1}{3}\sum_{j=x,y,z}\left(\frac{\dot{b}_j\ddot{b}_j}{\om_{j,0}^2}-\frac{\dot{b}_j}{b_j\Gamma^{2/3}}\right)\la\hat{H}(0)\ra\\
&=&\frac{d}{dt}\left(\frac{1}{2 \Gamma^{2/3}}+\frac{1}{6}\sum_{j=x,y,z}\frac{\dot{b}_j^2}{\om_{j,0}^2}\right)\la\hat{H}(0)\ra.
\eeqa
As for the variation of the expectation value of the trapping potential term, it simply reads
\beqa
\frac{d}{dt}\la \hat{H}_{\rm trap}(t)\ra&=&\frac{d}{dt}\left(\frac{1}{6}\sum_{j=x,y,z}\frac{\om_j^2(t)}{\om_{j,0}^2}b_j^2\right)\la\hat{H}(0)\ra.
\eeqa
The exact expression for the nonadiabatic mean energy directly follows from integrating the differential equation
\beqa
\frac{d}{dt}\la \hat{H}\ra=\frac{d}{dt}\la \hat{H}_0+\hat{H}_{\rm trap}(t)\ra,
\eeqa
with the boundary condition $\la\hat{H}(t=0)\ra=\la\hat{H}(0)\ra$ and reads
\beqa
\label{meanen}
\la\hat{H}(t)\ra=\left[\frac{1}{2\Gamma^{2/3}}+\frac{1}{6}\sum_{j=x,y,z}\frac{\dot{b}_j^2+\om_j^2(t)b_j^2}{\om_{j,0}^2}\right]\la\hat{H}(0)\ra,
\eeqa
which is the central result of this section, from which all  finite-time thermodynamics can be derived.

In the adiabatic limit under slow driving, the scaling factors remain coupled and fulfill
\beqa
b_{j,ad}(t)=\frac{1}{\Gamma_{ad,t}^{1/3}}\frac{\om_{j,0}}{\om_{j}(t)},\quad {\rm where } \quad
\Gamma_{ad}(t)=\prod_jb_{j,ad}(t).
\eeqa
Therefore, the instantaneous adiabatic mean energy reads
\beqa
\la\hat{H}(t)\ra_{ad}=\frac{1}{\Gamma_{ad}^{2/3}}\la\hat{H}(0)\ra.
\eeqa

The ratio between the nonadiabatic and adiabatic mean energies plays a crucial role in finite-time thermodynamics and is known as the
 the nonadiabatic factor \cite{Husimi53,Jaramillo16}
\begin{align}
Q^{\ast}(t)=\langle\hat{H}(t)\rangle/\langle\hat{H}(t)\rangle_{ad},
\end{align}
that quantifies quantum friction. For a unitary Fermi gas, the nonadiabatic factor and mean work read
\begin{align}
Q^*(t)&=\Gamma_{ad}^{2/3}\left[\frac{1}{2\Gamma^{2/3}}+\frac{1}{6}\sum_{j=x,y,z}\frac{\dot{b}_j^2+\om_j^2(t)b_j^2}{\om_{j,0}^2}\right],\\
\la W(t)\ra&=\la H(t)\ra-\la H(0)\ra=\left[\frac{1}{2\Gamma^{2/3}(t)}+\frac{1}{6}\sum_{j=x,y,z}\frac{\dot{b}_j^2+\om_j^2(t)b_j^2}{\om_{j,0}^2}-1\right]\la\hat{H}(0)\ra.
\end{align}

\subsection{Local counterdiabatic control of the finite-time thermodynamics of a unitary Fermi gas}
An arbitrary modulation in time of the trapping potential  generally induces a nonadiabatic evolution. We aim at tailoring excitations to control the finite-time thermodynamics of the system. This goal is at reach using techniques known as shortcuts to adiabaticity. Specifically, the counterdiabatic driving technique \cite{demirplak03,Berry09} can be exploited in combination with dynamical scaling laws to control the nonadiabatic dynamics of many-body quantum systems as discussed in \cite{delcampo11,delcampo13,DJD14}.

Our approach to achieve superadiabatic control relies on first designing a desirable reference  trajectory by specifying the modulation of the parameters of the Hamiltonian, i.e., $\om_j(t)$.
Subsequently, we consider an alternative protocol with frequencies $\om_j(t)\rightarrow \Om_j(t)$ that lead to the same final state than the adiabatic evolution corresponding to $\om_j(t)$. The modified protocol with trap frequencies $ \Om_j(t)$ can be found for an arbitrary value $t_f$ of the duration of the stroke, provided that $ \Om_j(t)$ can be implemented.


To design the reference evolution of the cloud, let $\{\omega_{j,0}|j=x,y,z\}$ denote the frequencies of the anisotropic harmonic trap at $t=0$. Similarly, let $\{b_{j,f}|j=x,y,z\}$ denote the target scaling factors upon completion of an expansion or compression
stroke of duration $t_f$. Imposing the following boundary conditions
\begin{eqnarray}
\omega_j(0)&=&\omega_{j,0}, ~~~\omega_j(t_f)=b_{j,f}^{-2}\omega_j(0),\\
\dot{\omega}_{j,0}&=&0,~~~~~~\dot{\omega}_{j,f}=0,\\
\ddot{\omega}_{j,0}&=&0,~~~~~~\ddot{\omega}_{j,f}=0,
\end{eqnarray}
we determine a possible choice for the time-dependent scaling factors via the polynomial ansatze $\omega_j(t)=\omega_{j,0}\sum_n c_n(t/t_f)^n$. Specifically,
\begin{equation}
\omega_j(t)=\omega_{j,0}\left[1+10\left(b_{j,f}^{-2}-1\right)\left(\frac{t}{t_f}\right)^3-15\left(b_{j,f}^{-2}-1\right)\left(\frac{t}{t_f}\right)^4+6\left(b_{j,f}^{-2}-1\right)\left(\frac{t}{t_f}\right)^5\right].
\end{equation}
Consistently, the reference expansion factor is set by the adiabatic consistency equations
\begin{align}
\label{bjadiab}
b_j(t)&=\frac{\omega_{j,0}}{\omega_j(t)\Gamma_{ad}^{1/3}(t)}\\
&=\frac{\omega_{j,0}}{\omega_{j,t}}\left[\frac{\nu(t)}{\nu(0)}\right]^{1/2},
\end{align}
where the geometric mean frequency equals $\nu(t)=[\omega_x(t)\omega_y(t)\omega_z(t)]^{1/3}$.

The above equations describe the evolution of the scaling factors in the adiabatic limit under slow driving. Nonetheless, they describe  as well the exact nonadiabatic dynamics under a modified driving protocol with a different time-dependence of the trapping frequencies, i.e., replacing $\omega_j(t)\rightarrow\Omega_j(t)$, where the explicit form of $\Omega_j(t)$ is to be determined. This approach, generally referred to as local counterdiabatic driving (LCD),  has been discussed for the single-particle time-dependent harmonic oscillator and many-body quantum systems in one spatial dimension and under isotropic confinement in higher dimensions \cite{delcampo11,delcampo13,DJD14}. For a unitary Fermi gas in a 3D anisotropic trap,  the requiring driving frequencies are given by
\begin{align}
\Omega_j^2(t)&=\frac{\omega_{j,0}^2}{b_j^2\Gamma^{2/3}}-\frac{\ddot{b}_j}{b_j}\\
&=\omega_j^2(t)-\frac{\ddot{b}_j}{b_j}\ ,
\end{align}
where $b_j(t)$ are given by Eqs. (\ref{bjadiab}).
The explicit expression for $\Om_j^2(t)$ reads
\beqa
\Om_j^2(t)&=&\om_j^2-2\left(\frac{\dot{\om}_j}{\om_j}\right)^2+\frac{\ddot{\om}_j}{\om_j}-\frac{4}{9}\left(\frac{\dot{\Gamma}}{\Gamma}\right)^2+3\frac{\ddot{\Gamma}}{\Gamma}-\frac{2}{3}\frac{\dot{\om}_j\dot{\Gamma}}{\om_j\Gamma}\\
&=&\om_j^2-2\left(\frac{\dot{\om}_j}{\om_j}\right)^2+\frac{\ddot{\om}_j}{\om_j}
+\frac{3}{4}\left(\frac{\dot{\nu}}{\nu}\right)^2-\frac{1}{2}\frac{\ddot{\nu}}{\nu}
+\frac{\dot{\om}_j\dot{\nu}}{\om_j\nu}\, , \label{eq:Omega}
\eeqa
which includes counterdiabatic corrections arising from $\om_j$, the geometric mean $\nu$, and their coupling. The nonadiabatic factor for a unitary Fermi gas under local counterdidabatic driving reads
\beqa
Q^*(t)&=&1+\frac{1}{6}\Gamma(t)^{2/3}\sum_j\frac{\dot{b}_j^2-b_j\ddot{b}_j}{\om_{j,0}^2}\\
&=&1+\frac{1}{6}\sum_j\left(\frac{\ddot{\om}_j}{\om_j^3}-\frac{\dot{\om}_j^2}{\om_j^4}-\frac{\ddot{\nu}}{2\nu\om_j^2}+\frac{\dot{\nu}^2}{2\nu^2\om_j^2}\right).
\eeqa
When the scaling factors are isotropic so that $b_j(t)= b(t)$, the axis subindex $j$ can be dropped out, the volume scaling factor
simplifies to $\Gamma(t)=b(t)^3=[\om_0/\om(t)]^{\frac 3 2}$ and
\begin{align}
\Om^2(t)=\om^2(t)-\frac{3}{4}\left[\frac{\dot{\om}(t)}{\om(t)}\right]^2+\frac{1}{2}\frac{\ddot{\om}(t)}{\om(t)}\ ,
\end{align}
in agreement with ~\cite{delcampo13,DJD14}. Similarly, in the isotropic case the nonadiabatic factor reduces to
\begin{equation}
Q^*(t)=1+\frac{1}{4}\left(\frac{\ddot{\om}}{\om^3}-\frac{\dot{\om}^2}{\om^4}\right),
\end{equation}
which agrees for the result in one-dimensional harmonically trapped systems~\cite{Beau16,Abah17}.

\section{Experiment methods}

The experimental setup is  similar to that described in our previous work~\cite{Wu1,Wu2}. We realize a large atom-number magneto-optical trap (MOT) by employing a laser system of 2.5-watts laser output with Raman fiber amplifier and intracavity frequency-doubler. When the MOT loading and cooling stage are finished, an optical pumping process is performed and a balanced mixture of atoms in the two lowest hyperfine states, $|\!\!\uparrow\rangle\equiv|F=1/2, M_F=-1/2\rangle$ and $|\!\!\downarrow\rangle\equiv|F=1/2, M_F= 1/2\rangle$, is prepared.
\begin{figure}[htb]
\centering
\includegraphics[width=16cm]{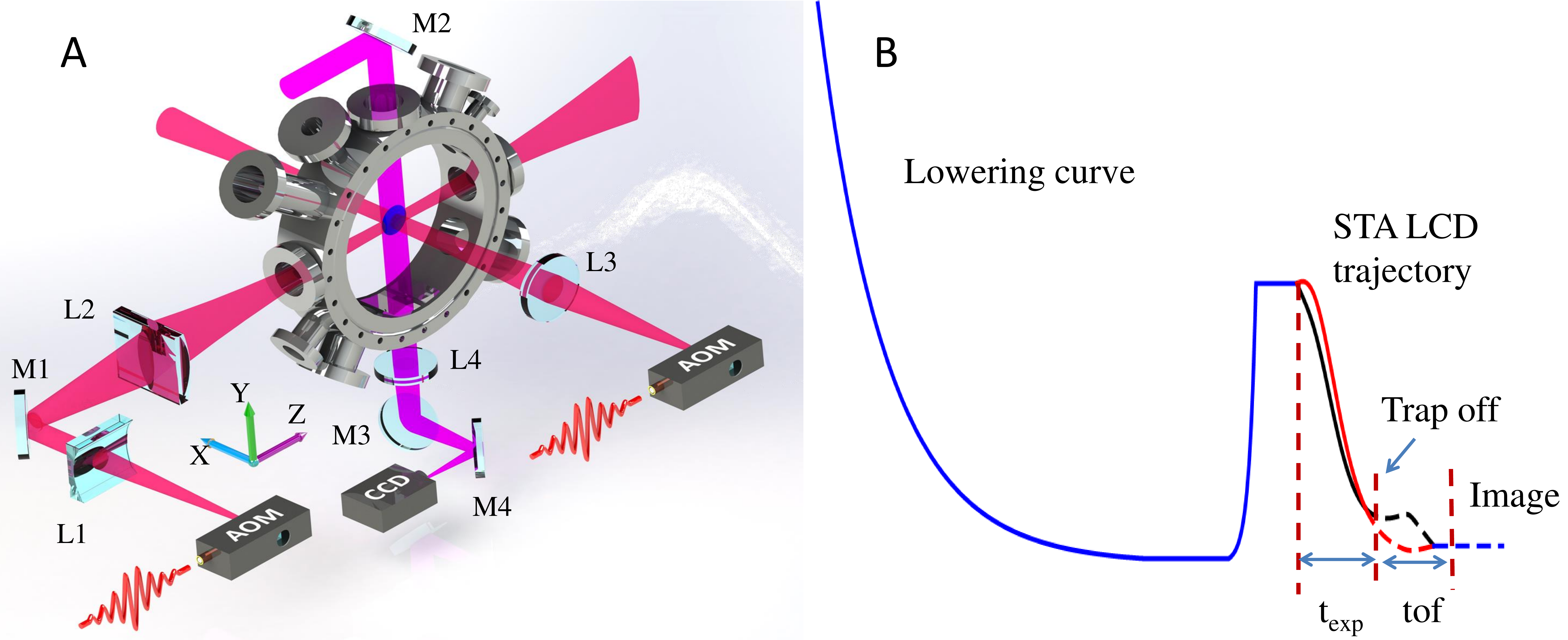}
\caption{Schematic of the experimental setup (A) and the experimental procedure (B). A specially designed optical crossed-dipole trap is formed by two orthogonal far-off resonance laser beams, providing a highly controllable trap frequency. M1-M4, Mirrors; L1-L2, cylindrical lenes; L3-L4, achromatic lenses; AOM, acousto-optic modulator; tof, time-of-flight.  }\label{figsetup}
\end{figure}

After the MOT stage, the atoms are first loaded into an optical dipole trap formed by a single beam. A forced evaporation is performed to cool atoms to quantum degeneracy in an external magnetic field at 832 G. Two seconds later a specially designed optical crossed dipole trap is switched on and nearly $10^5$ atoms are transferred to this trap. Further forced evaporation is implemented by slightly reducing the power of the two orthogonal beams. In this way, the system is initially prepared   in a stationary state with a dimensionless temperature $T/T_F\approx0.25$, where $T_F$ is the Fermi temperature. Subsequently, we start the STA experiment based on LCD, schematically described in Fig. \ref{figsetup}B. After the evolved time $t_{exp}$, the beams of the crossed dipole trap are switched off and an absorption image is taken with the time of flight.

The specially-designed optical dipole trap, which consists of two orthogonal far-off resonance laser beams, provides a flexible control of the trap frequencies, as is shown in Fig. \ref{figsetup}A. The first beam propagates along the  $z$-axis, i.e., the horizontal direction. The beam is focused by a pair of cylindrical lens, which cause the more tight confinement in the $x$ direction. The Gaussian waists are 42 $\mu$m along the $x$-axis and about 1200 $\mu$m in the $y$ direction. The second beam, perpendicular to the first,  propagates along the $x$ direction, i.e. the vertical beam. The beam has a nearly perfect Gaussian profile, with a waist  of $\approx$ 68 $\mu$m, which provides both strong confinement in the $y$  and $z$ directions. Therefore, the frequency in the $x$ direction mostly depends on the power of the horizontal beam while the vertical beam determines the frequencies  both in the $y$ and $z$ directions. The frequency aspect ratio of the trap can be simply controlled by precisely adjusting the power ratio of the two beams.

\section{Expansion factor}
The dynamics of the system is investigated by measuring the radius of the atomic cloud. At different stages of the superadiabatic stroke, an absorption image is taken after releasing the atomic cloud in a time-of-flight expansion. The observed cloud size can be related to cloud size inside the trap,  by accounting for the time-of-flight expansion scaling factor $b_{j}$ along each axis ($j=x,y,z$). The time-of-flight density profile along each direction is fitted by a Gaussian function as $A_0+A_1e^{-r_j^2/\sigma_j^2}$ in the normal fluid regime. From this fit we obtain the observed cloud size $\langle\sigma_j\rangle_{obs}$. $\langle\sigma_j\rangle_{obs}$ is related to the in situ cloud size  via $\langle\sigma_j\rangle=\langle\sigma_j\rangle_{obs}/b_{j}$. Because $\sigma_j$ is obtained by a Gaussian fit to the density profile, $\langle\sigma_j^2\rangle=2\langle R_j^2\rangle$, and thus the expansion dynamics can be monitored by directly measuring the density profile.
\begin{figure}[htb]
\centering
\includegraphics[width=8cm]{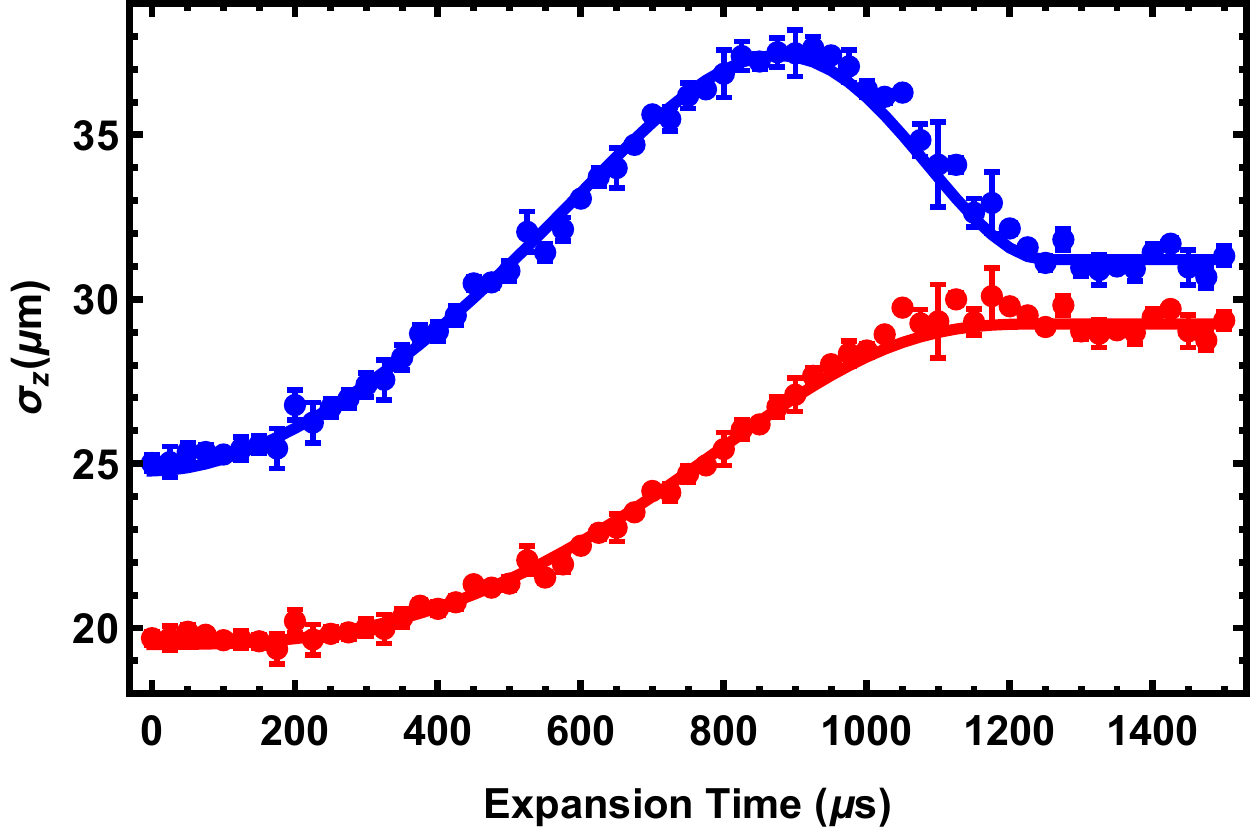}
\caption{Determination of the cloud size $\langle\sigma_z\rangle$ obtained from the Gaussian fit of the density profile of the unitary Fermi gas for a time-of-flight  $t_{tof} = 600\,\mu$s. The blue dots are the data $\langle\sigma_z\rangle_{obs}$ and the red dots are $\langle\sigma_z\rangle=\langle\sigma_z\rangle_{obs}/b_{tof}$. The red solid line is the best fit based on theory curve. The blue solid line is the red solid line multiplied by $b_{tof}$. Error bars represent the standard deviation of the statistics.}\label{fig2}
\end{figure}

To analyze the measurements we use the equations of motion for the scaling factors of a unitary Fermi gas
\begin{equation}
\ddot{b}_j+\omega_j^2(t)b_j=\frac{\omega_{j,0}^2}{b_j\Gamma^{2/3}}\ ,~~~j=x,y,z ~\label{eq12}
\end{equation}
where  $b_j$ denotes the expansion scaling factor in $j$ direction and $\omega_{j,0}$ is the initial frequency measured by the parametric resonance for the atomic cloud. During  the local counterdiabatic driving of the system, the time-dependent frequencies are given by
\begin{equation}\label{eq13}
\omega_j(t)=\left\{
\begin{aligned}
\Omega_j(t),~~~t_0<&t\leq t'\\
0,~~~t'<&t\leq t'+t_{tof}
\end{aligned}
\right.
\end{equation}
Here $t_0$ is the time when we start the STA process, $t'= t_0+t_{exp}$ is the time at which we turn off the trap, and $t_{tof}$ is the duration of the time-of-flight expansion for imaging of the cloud. Using equations (\ref{eq12}) and (\ref{eq13}), we can numerically calculate the expansion factors at any time $t$ using the boundary conditions $b_j(t_0)=1$ and $\dot{b}_j(t_0)=0$ for all $j=x,y,z$. In Fig. \ref{fig2}, we show $\langle\sigma_z\rangle$ and $\langle\sigma_z\rangle_{obs}$ of the unitary Fermi gas for a time-of-flight $t_{tof} = 600\mu$s.

\section{Expansion and compression stroke}

A quantum Otto cycle  consists of a sequence of four strokes, with  isentropic and isochoric strokes alternating with each other \cite{KR17,Beau16}. We focus on  the implementation of the isentropic expansion and the  compression strokes during which the dynamics is unitary. Utilizing STA based on LCD driving, we implement the superadiabatic variants of the expansion and compression strokes with a 3D unitary Fermi gas as a working medium. Experimentally, we consider  two different trap configurations, associated with the end points of the expansion and compression stages. The corresponding  trap frequencies are
\begin{align}
\Phi_1&: ~~~~~~\omega_{x1}=2\pi\times1600~\text{Hz},~~~~\omega_{y1}=\omega_{z1}=2\pi\times200~\text{Hz},\\
\Phi_2&: ~~~~~~\omega_{x2}=2\pi\times~400~~\text{Hz},~~~~\omega_{y2}=\omega_{z2}=2\pi\times200~\text{Hz}.
\end{align}
The energies of the corresponding quantum states $\Phi_1$ and $\Phi_2$, are denoted  by  $\langle H_1\rangle$ and $\langle H_2\rangle$, respectively.
The expansion stroke is performed from state $\Phi_1$ to state $\Phi_2$ and the compression stroke is manipulated by reversing the expansion procedure. The frequencies of the two strokes have the following relations,
\begin{align}
\Omega_{j,exp}(0)&=\omega_{j,1},~~~~~\Omega_{j,exp}(t_f)=\omega_{j,2},~~~~~\Omega_{j,exp}(t)=\Omega_{j,1}(t),\label{eq4}\\
\Omega_{j,comp}(0)&=\omega_{j,2},~~~\Omega_{j,comp}(t_f)=\omega_{j,1},~~~\Omega_{j,comp}(t)=\Omega_{j,1}(t_f-t), \label{eq5}
\end{align}
where $\Omega_{j,exp}$ and $\Omega_{j,comp}$ are the frequencies for the expansion and compression strokes, respectively.

Along the expansion stroke, the aspect ratio is initially fixed to be 8 and finally reaches the value 2 at time $t_f$. The reverse processes associated with the compression stroke thus changes the trap frequency aspect ratio from 2 to 8.

\begin{figure}[htb]
\begin{minipage}{0.4\textwidth}
\includegraphics[width=\textwidth,height=1.9 in]{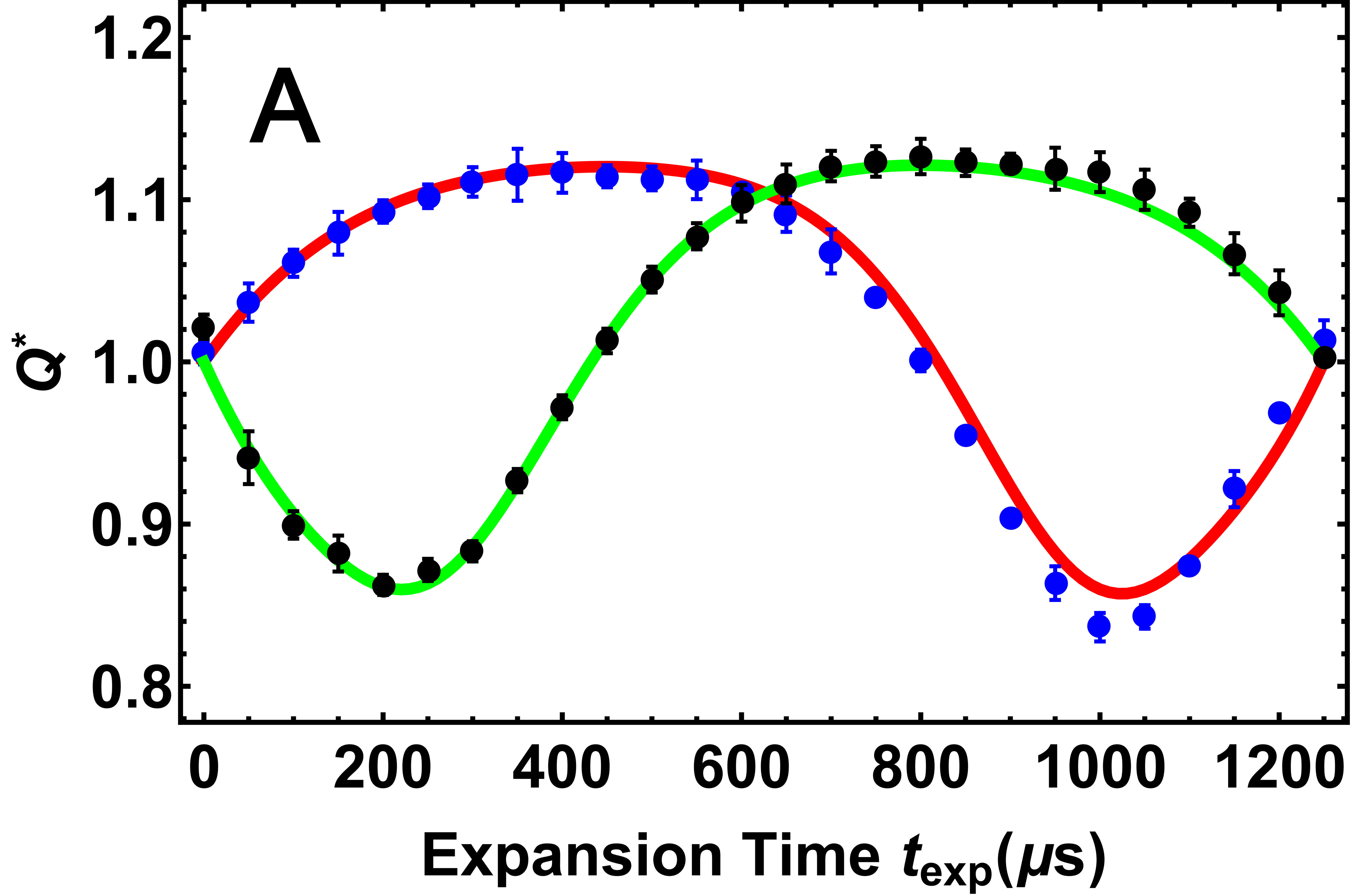}
\end{minipage}
\hspace{0.1\textwidth}
\begin{minipage}{0.4\textwidth}
\includegraphics[width=\textwidth,height=1.9 in]{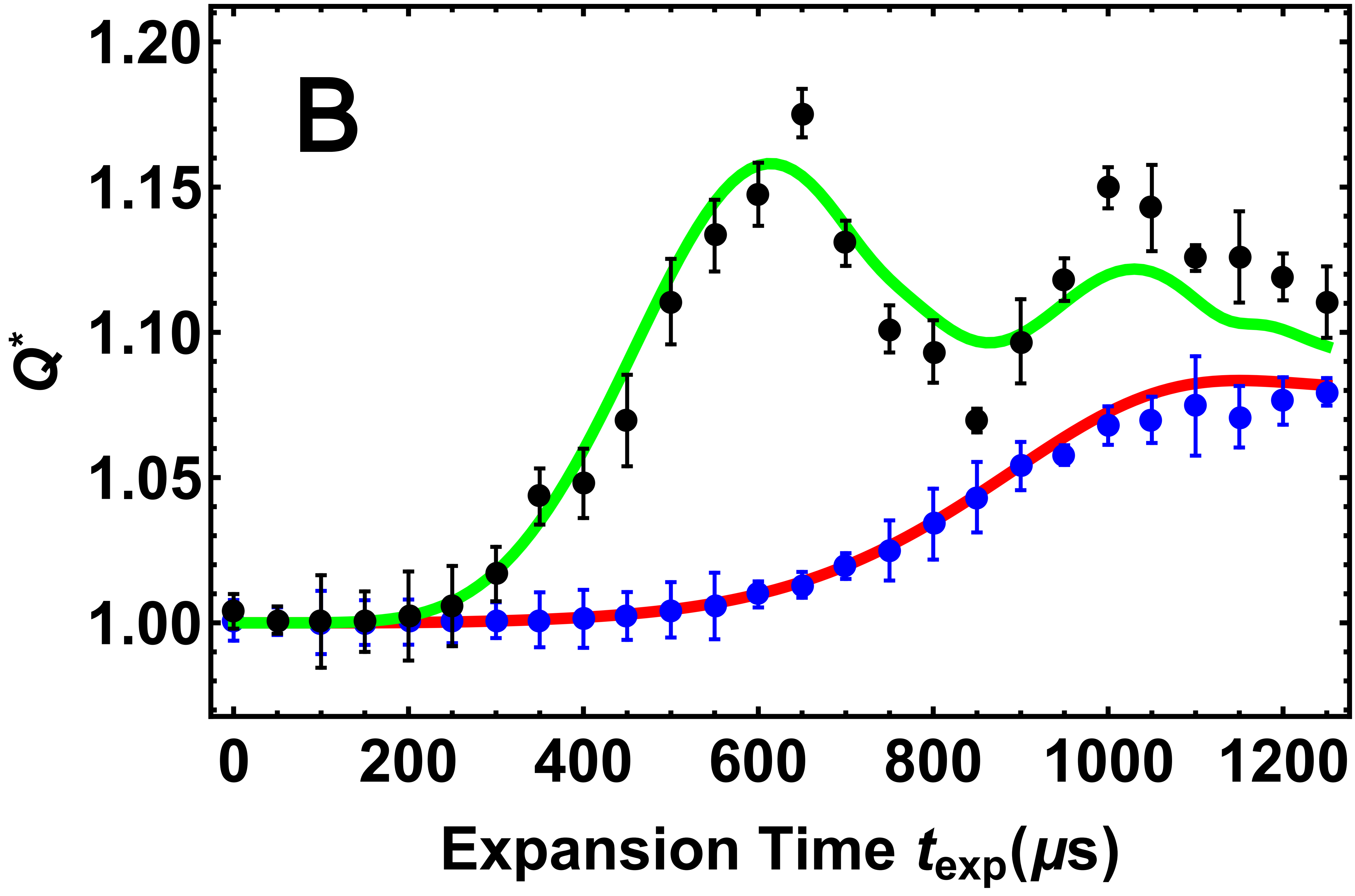}
\end{minipage}
\caption{The evolution of the  non-adiabatic factor $Q^*(t)$ for  the expansion  and compression strokes. (A) $Q^*(t)$ for STA with LCD driving. (B) $Q^*(t)$ for the reference driving. For both (A) and (B), blue dots and black dots represent measurement data for the expansion and compression strokes, respectively, while the red line and green line are  the corresponding theoretical predictions. Error bars represent the standard deviation of the measurement statistics.}\label{fig3}
\end{figure}

The measured results for the  non-adiabatic factor $Q^*(t)$ are shown in  Fig.~\ref{fig3}.  The non-adiabatic factor of both the expansion and compression strokes for LCD driving go back to 1 at the time $t=t_f$, showing that there are frictionless, i.e., satisfying $Q^*(t_f)=1$. By contrast,  the reference driving exhibits  friction with values of  $Q^*(t_f)$ greater than unity. Therefore, we demonstrate that  a STA protocol by LCD driving of a unitary Fermi gas is not only suitable for an expansion stroke, but also for its superadiabatic  compression in finite time. Based on this fast driving scheme, one can induce frictionless control so as to transfer strongly interacting fluids between thermal equilibrium states.

Note that while in a STA energy excitations are transient and they vanish upon completion of the stroke, there are large energy excitations for the compression stroke of the nonadiabatic reference driving process in the final state $t=t_f$. The shear viscosity  can not be neglected at the high energy of unitary Fermi gas~\cite{Thomas1,Thomas2} and we take it into consideration for the theoretical calculations of the reference driving of the compression stroke in Fig.~\ref{fig3}.

The mean work for the LCD driving is determined by the geometric frequency according to equation (\ref{eq1}). The theoretical  mean work  is therefore
\begin{align}
\langle W_{exp}(t_f)\rangle&=(1/\Gamma_{exp,ad}^{2/3}(t_f)-1)\langle H(0)\rangle_1,\\
\langle W_{comp}(t_f)\rangle&=(1/\Gamma_{comp,ad}^{2/3}(t_f)-1)\langle H(0)\rangle_2,
\end{align}
where $\langle H(0)\rangle_1$ and $\langle H(0)\rangle_2$ are the equilibrium mean energies of the states $ \Phi_1$ and $ \Phi_2$, respectively. Given that the adiabaticity condition is fulfilled upon completion of the adiabatic and superadiabatic strokes, one  finds that
\begin{align}
\langle W_{exp}(t_f)\rangle+W_{comp}(t_f)=0,\\
\Gamma_{exp,ad}(t_f)=1/\Gamma_{comp,ad}(t_f).
\end{align}
 To characterize the evolution of the mean work, we define the dimensionless factor $\kappa$ as
\begin{equation}
\kappa=\langle W(t)\rangle/\langle H(0)\rangle.
\end{equation}

 Fig. \ref{fig4} shows the evolution of $\kappa$ for both the expansion  and compression strokes. The energy of the evolving state in the expansion process decreases and is below its initial vale, therefore $\kappa$ acquires negative values. However, the energy in the compression process is increased and  $\kappa$ is positive. Upon completion of the stroke, the mean work  for the reference driving is always larger than the one for the corresponding LCD driving, indicating that an excess energy is added for nonadiabatic driving over the adiabatic value. This excess of energy results from the  existence of friction for the reference driving. In a thermodynamic cycle,  friction in the strongly interacting quantum fluids would generate excess excitations, reducing the efficiency of the output power for the heating engine.

\begin{figure}[htb]
\centering
\includegraphics[width=10cm]{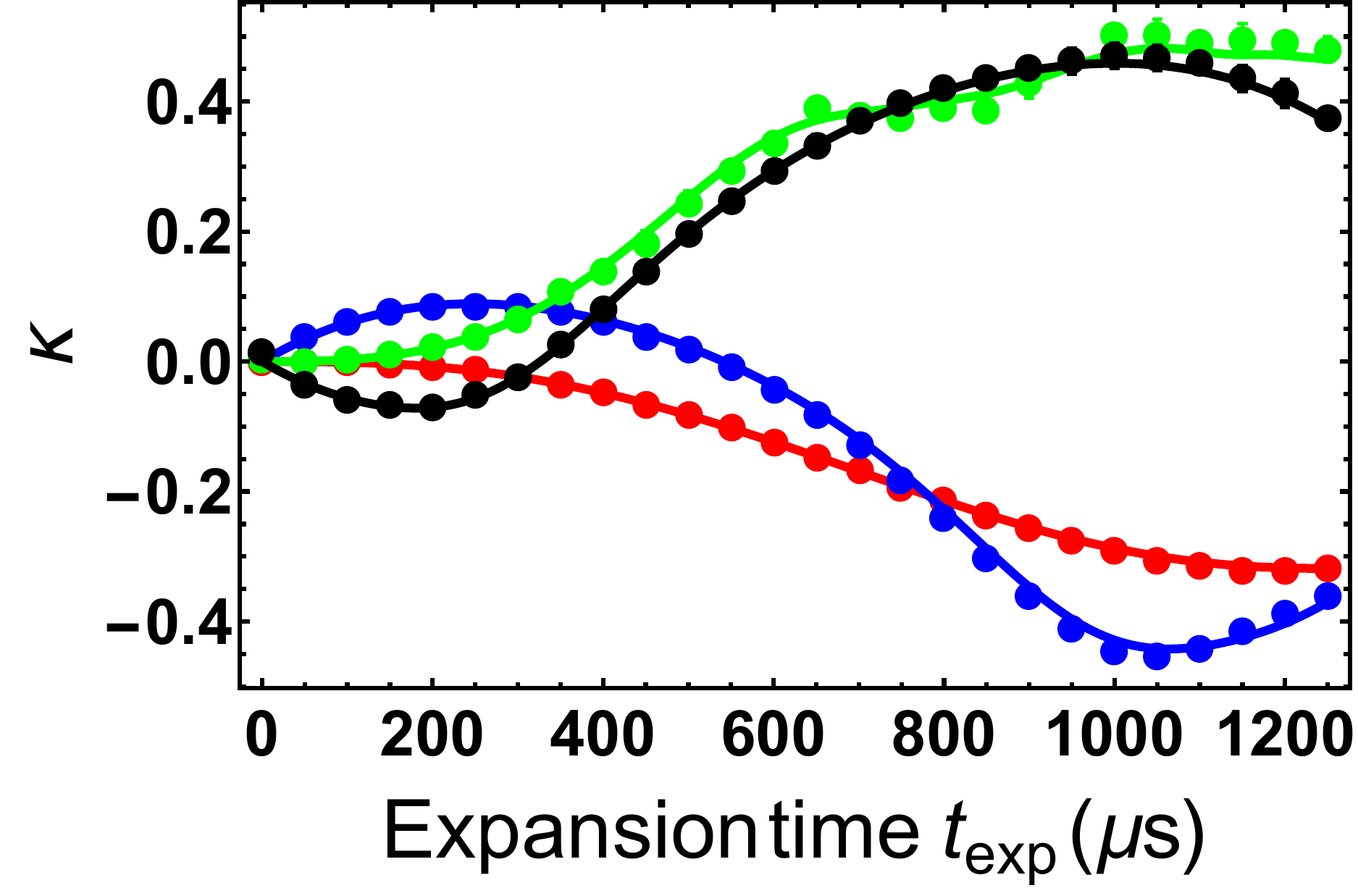}
\caption{Mean work $\kappa=\langle W(t)\rangle/\langle H(0)\rangle$, for both the expansion stroke and compression stroke. Blue  and red dots represent measured data for the expansion stroke while black  and green dots are measured results for the compression stroke. Solid lines correspond to the theoretical predictions of the LCD and reference protocols. Error bars represent the standard deviation of the statistics.}
\label{fig4}
\end{figure}

In the following, we consider the adiabatic ``slow'' driving case, where the quantum states are stationary at any time and can be reversed. The mean work for the the expansion and compression strokes can be written as,
\begin{align}
\int_0^tdW_{exp}(t)=-\int_{tf-t}^{t_f}dW_{comp}(t)&=-\langle W_{comp}(t_f)\rangle+\int_{0}^{t_f-t}dW_{comp}(t), \label{eq6} \\
\langle W_{exp}(t_f)\rangle&=-\langle W_{comp}(t_f)\rangle. \label{eq7}
\end{align}

Combining Eq. (\ref{eq6}) and Eq. (\ref{eq7}), we find
\begin{align}
\langle W_{exp}(t)\rangle-\langle W_{comp}(t_f-t)\rangle-\langle W_{exp}(t_f)\rangle=0.
\end{align}

\begin{figure}[htb]
\centering
\includegraphics[width=10cm]{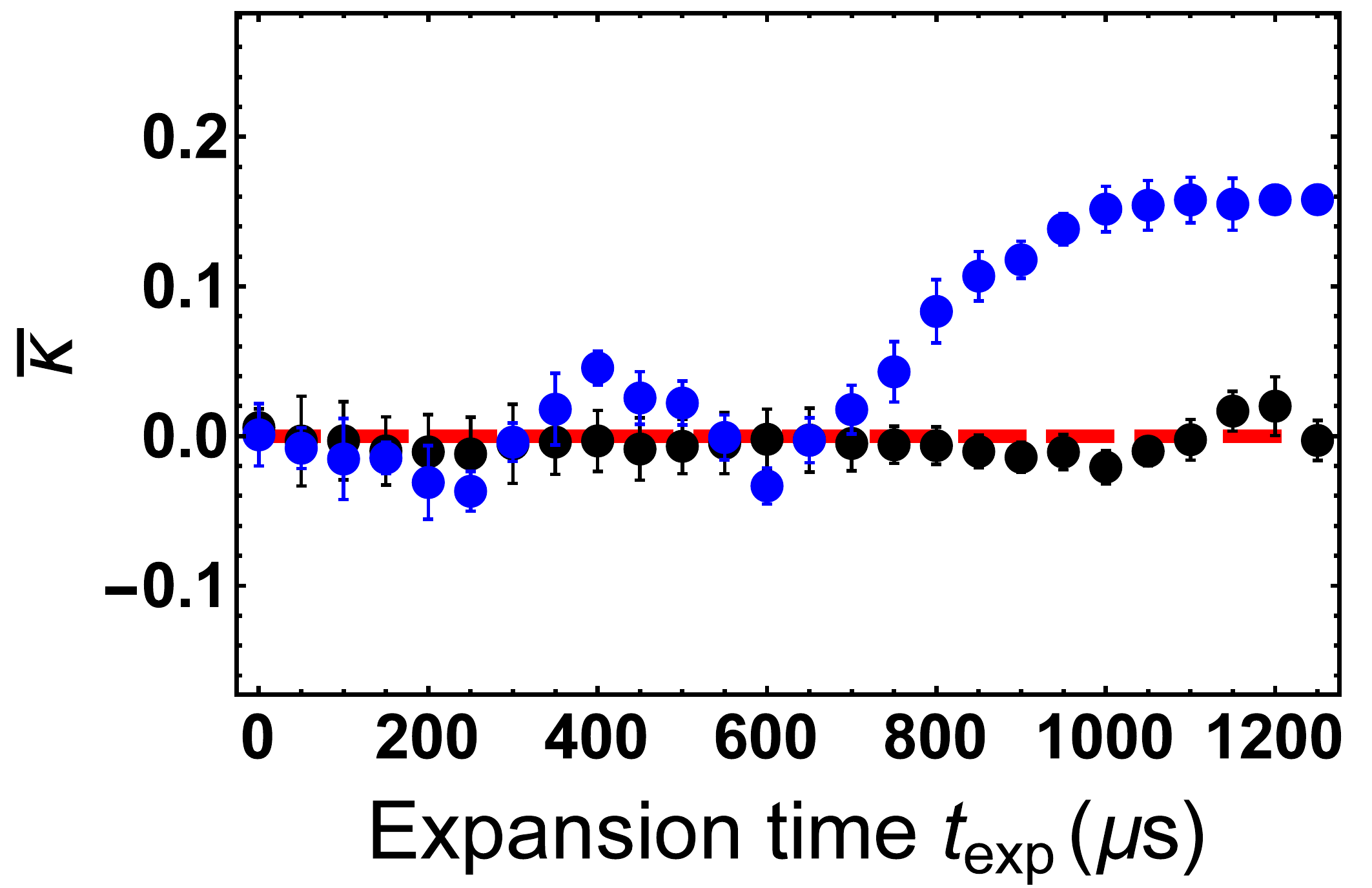}
\caption{Time evolution of the dimensionless factor $\bar{\kappa}$ for both the expansion  and compression strokes, $\bar{\kappa}=(\langle W_{exp}(t)\rangle-\langle W_{comp}(t_f-t)\rangle)/\langle H(0)\rangle_{1}-1$. Black  and blue dots correspond to the  measured data for LCD and reference driving, respectively. The red dashed line represents $\bar{\kappa}=0$. Error bars represent the standard deviation of the statistics.}\label{fig5}
\end{figure}

Here we define another dimensionless factor $\bar{\kappa}=(\langle W_{exp}(t)\rangle-\langle W_{comp}(t_f-t)\rangle)/\langle H(0)\rangle_{1}-1$, which denotes  the deviation from the adiabatic evolution. The factor $\bar{\kappa}$ should be zero at any time for the real adiabatic process. The measured results are presented in Fig. \ref{fig5}. It shows that $\bar{\kappa}$ for LCD driving is closed to zero, indicating the counter-adiabatic following of the adiabatic trajectory. By contrast,  for the reference driving  deviations from  zero value are clearly observed.

\begin{thebibliography}{99}


\bibitem{CA75} F. L. Curzon and B. Ahlborn, \href{http://dx.doi.org/10.1119/1.10023}{Am. J. Phys.  {\bf 43}, 22 (1975)}.


\bibitem{FTT84} B. Andresen, P. Salamon, R. S.  Berry,
\href{http://doi.org/10.1063/1.2916405}{ Phys. Today {\bf 37}, 62 (1984).}


\bibitem{Andresen} B. Andresen, \href{http://doi.org/10.1002/anie.201001411}{Angew. Chem. Int. Ed. 50, 2690 (2011).}


\bibitem{Alicki79} R. Alicki,
\href{10.1088/0305-4470/12/5/007/}{J. Phys A: Math. Gen., {\bf 12}, L103 (1979).}

\bibitem{Kosloff84} R. Kosloff,
 \href{http://dx.doi.org/10.1063/1.446862}{J. Chem. Phys.  {\bf 80}, 1625 (1984).}

\bibitem{Scully03}  M. O. Scully, M. S. Zubairy, G. S. Agarwal, and H. Walther, \href{https://doi.org/10.1126/science.1078955 }{Science {\bf 299}, 862 (2003).}


\bibitem{Rossnagel14} J . Ro\ss nagel, O. Abah, F. Schmidt-Kaler, K. Singer, and E. Lutz, \href{https://doi.org/10.1103/PhysRevLett.112.030602}{Phys. Rev. Lett. {\bf 112}, 030602  (2014).}

\bibitem{Rossnagel16} J. Ro\ss nagel, S. T. Dawkins, K. N. Tolazzi, O. Abah, E. Lutz, F. Schmidt-Kaler, K. Singer, \href{https://doi.org/10.1126/science.aad6320.}{Science {\bf 352}, 325 (2016).}

\bibitem{Maslennikov17}
G. Maslennikov, S. Ding, R. Hablutzel, J. Gan, A. Roulet, S. Nimmrichter, J. Dai, V. Scarani, D. Matsukevich, \href{https://arxiv.org/abs/1702.08672}{arXiv:1702.08672}

\bibitem{Kato50} T. Kato,  \href{https://doi.org/10.1143/JPSJ.5.435}{J. Phys. Soc. Jpn. {\bf 5}, 435 (1950).}

\bibitem{GK92} E. Geva and R. Kosloff,
\href{http://dx.doi.org/10.1063/1.463909}{J. Chem. Phys. {\bf 97}, 4398 (1992).}


\bibitem{YK06} Y. Rezek and R. Kosloff,
\href{http://dx.doi.org/10.1088/1367-2630/8/5/083}{New J. Phys. {\bf 8}, 83 (2006).}

\bibitem{Shiraishi16} N. Shiraishi,  K. Saito,  and H. Tasaki,  \href{https://doi.org/10.1103/PhysRevLett.117.190601}{Phys. Rev. Lett. {\bf 117}, 190601 (2016).}

    \bibitem{STAR} E. Torrontegui, S. Ib\'{a}\~{n}ez, S. Mart\'{i}nez-Garaot, M. Modugno, A. del Campo, D. Gu\'{e}ry-Odelin, A. Ruschhaupt, X. Chen, J. G. Muga,
\href{	10.1016/B978-0-12-408090-4.00002-5}{Adv. At. Mol. Opt. Phys. {\bf 62}, 117-169 (2013).}


\bibitem{Deng13} J. Deng, Q. Wang, Z. Liu, P. H\"anggi, and J. Gong,
\href{http://dx.doi.org/10.1103/PhysRevE.88.062122}{Phys. Rev. E {\bf 88}, 062122 (2013).}

\bibitem{delcampo14} A. del Campo, J. Goold, and M. Paternostro,
 \href{http://dx.doi.org/10.1038/srep06208}{Sci. Rep. {\bf 4}, 6208 (2014).}

\bibitem{Beau16} M. Beau, J. Jaramillo, A. del Campo,
 \href{http://dx.doi.org/10.3390/e18050168}{Entropy {\bf 18}, 168 (2016).}


\bibitem{Abah16} O. Abah and  E. Lutz,
\href{https://doi.org/10.1209/0295-5075/118/40005}{EPL {\bf 118}, 40005 (2017)}.

 \bibitem{demirplak03} M. Demirplak, and Stuart A. Rice,
 \href{http://dx.doi.org/10.1021/jp030708a}{J. Phys. Chem. A {\bf 107}, 9937 (2013).} 

\bibitem{Berry09} M. V. Berry,
\href{http://dx.doi.org/10.1088/1751-8113/42/36/365303}{J. Phys. A: Math. Theor. {\bf 42}, 365303 (2009).} 

  \bibitem{expCD1} M. G. Bason, M. Viteau, N. Malossi, P. Huillery, E. Arimondo, D. Ciampini, R. Fazio, V. Giovannetti, R.  Mannella, O. Morsch,  \href{http://dx.doi.org/10.1038/nphys2170}{Nature Phys. {\bf 8}, 147 (2012).}

\bibitem{expCD2} J. Zhang, J. Hyun Shim, I. Niemeyer, T. Taniguchi, T. Teraji, H. Abe, S. Onoda, T. Yamamoto, T. Ohshima, J. Isoya, D. Suter,
\href{http://dx.doi.org/10.1103/PhysRevLett.110.240501}{Phys. Rev. Lett. {\bf 110}, 240501 (2013).}

\bibitem{expCD3} Yan-Xiong Du, Zhen-Tao Liang, Yi-Chao Li, Xian-Xian Yue, Qing-Xian Lv, Wei Huang, Xi Chen, Hui Yan, Shi-Liang Zhum, \href{http://dx.doi.org/10.1038/ncomms12479}{Nat. Commun. {\bf 7}, 12479 (2016). }

\bibitem{expCD4} Shuoming An, Dingshun Lv, Adolfo del Campo, Kihwan Kim,  \href{http://dx.doi.org/10.1038/ncomms12999}{Nat. Commun. {\bf 7}, 12999 (2016).}

 \bibitem{DTSon} Y. Nishida and D. Son, \href{https://doi.org/10.1103/PhysRevD.76.086004}{Phys. Rev. D{\bf 76}, 086004 (2007).}

\bibitem{BCSBEC} W. Zwerger, The BCS-BEC Crossover and the Unitary Fermi Gas, (Springer Berlin Heidelberg, 2012)

  \bibitem{Talkner07}
P. Talkner, E. Lutz, P. Hanggi,  \href{https://doi.org/10.1103/PhysRevE.75.050102}{Phys. Rev. E {\bf 75}, 050102(R) (2007).}

 \bibitem{anisotropicexpansion}
K. M. O'Hara, S. L. Hemmer, M. E. Gehm, S. R. Granade, J. E. Thomas, \href{https://doi.org/10.1126/science.1079107}{Science {\bf 298}, 2179 (2002).}

\bibitem{3DFermi}
C. Menotti, P. Pedri, and S. Stringari, \href{https://doi.org/10.1103/PhysRevLett.89.250402}{Phys. Rev. Lett. {\bf 89}, 250402 (2002).}


\bibitem{Husimi53} K. Husimi,
\href{http://dx.doi.org/10.1143/ptp/9.4.381}{\textit{Prog. Theor. Phys.} {\bf 9,} 381 (1953).}
\bibitem{Jaramillo16}
J. Jaramillo, M. Beau, A. del Campo,
  \href{http://dx.doi.org/10.1088/1367-2630/18/7/075019}{New J. Phys. {\bf 18}, 075019 (2016).}

\bibitem{delcampo13} A. del Campo,
\href{http://dx.doi.org/10.1103/PhysRevLett.111.100502}{Phys. Rev. Lett. {\bf 111}, 100502 (2013).} 

\bibitem{DJD14} S. Deffner, C. Jarzynski, and A. del Campo,
\href{http://dx.doi.org/10.1103/PhysRevX.4.021013}{Phys. Rev. X {\bf 4}, 021013 (2014).}




\bibitem{Wu1}
S. Deng, P. Diao, Q. Yu, and H. Wu,  \href{https://doi.org/10.1088/0256-307X/32/5/053401}{Chin. Phys. Lett. {\bf 32}, 053401 (2015).}

\bibitem{Wu2}
S. Deng, Z. Shi, P. Diao, Q. Yu, H. Zhai, R. Qi, and H. Wu, \href{https://doi.org/10.1126/science.aaf0666 }{Science {\bf 353}, 371 (2016).}







\bibitem{Salamon09} P. Salamon, K. H. Hoffmann, Y. Rezek, and R. Kosloff,
\href{http://dx.doi.org/10.1039/B816102J }{Phys. Chem. Chem. Phys. {\bf 11}, 1027 (2009).}

\bibitem{Rezek09} Y. Rezek, P. Salamon, K. H. Hoffmann, and R. Kosloff,
\href{http://dx.doi.org/10.1209/0295-5075/85/30008}{EPL {\bf 85}, 30008 (2009).}

\bibitem{Stefanatos10} D. Stefanatos, J. Ruths, and J.-S. Li, 
 \href{http://dx.doi.org/10.1103/PhysRevA.82.063422}{Phys. Rev. A {\bf 82}, 063422 (2010).}

\bibitem{Hoffmann11} K. H. Hoffmann, P. Salamon, Y. Rezek, and R. Kosloff,
\href{	http://dx.doi.org/10.1209/0295-5075/96/60015}{EPL {\bf 96}, 60015 (2011).}







\end{thebibliography}

\begin{thebibliography}{100}
\bibitem{Husimi53} K. Husimi,
\href{http://dx.doi.org/10.1143/ptp/9.4.381}{Prog. Theor. Phys. {\bf 9,} 381 (1953).}

\bibitem{Jaramillo16}
J. Jaramillo, M. Beau, A. del Campo,
 \href{http://dx.doi.org/10.1088/1367-2630/18/7/075019}{New J. Phys. {\bf 18}, 075019 (2016).}

\bibitem{demirplak03} M. Demirplak, and Stuart A. Rice,
\href{http://dx.doi.org/10.1021/jp030708a}{J. Phys. Chem. A {\bf 107}, 9937 (2013).} 

\bibitem{Berry09} M. V. Berry,
\href{http://dx.doi.org/10.1088/1751-8113/42/36/365303}{J. Phys. A: Math. Theor. {\bf 42}, 365303 (2009).} 



\bibitem{delcampo11} A. del Campo,
\href{http://dx.doi.org/10.1209/0295-5075/96/60005}{EPL {\bf 96}, 60005 (2011). }

\bibitem{delcampo13} A. del Campo,
\href{http://dx.doi.org/10.1103/PhysRevLett.111.100502}{Phys. Rev. Lett. {\bf 111}, 100502 (2013).} 

\bibitem{DJD14} S. Deffner, C. Jarzynski, and A. del Campo,
 \href{http://dx.doi.org/10.1103/PhysRevX.4.021013}{Phys. Rev. X {\bf 4}, 021013 (2014).}

\bibitem{Beau16} M. Beau, J. Jaramillo, A. del Campo,
 \href{http://dx.doi.org/10.3390/e18050168}{Entropy {\bf 18}, 168 (2016).}


\bibitem{Abah17} O. Abah and  E. Lutz,
\href{https://doi.org/10.1209/0295-5075/118/40005}{EPL {\bf 118}, 40005 (2017)}.



\bibitem{Wu1}
S. Deng, P. Diao, Q. Yu, and H. Wu,  \href{https://doi.org/10.1088/0256-307X/32/5/053401}{Chin. Phys. Lett. {\bf 32}, 053401 (2015).}

\bibitem{Wu2}
S. Deng, Z. Shi, P. Diao, Q. Yu, H. Zhai, R. Qi, and H. Wu, \href{https://doi.org/10.1126/science.aaf0666 }{Science {\bf 353}, 371 (2016).}


\bibitem{KR17} R. Kosloff and  Y.  Rezek, \href{10.3390/e19040136}{Entropy  {\bf 19}, 136 (2017).}


\bibitem{Thomas1}
C. Cao, E. Elliott, J. Joseph, H. Wu, J. Petricka, T. Sch\"{a}fer, J. E. Thomas, \href{https://doi.org/10.1126/science.1195219}{Science {\bf 331}, 58 (2011).}

\bibitem{Thomas2} E. Elliott, J. A. Joseph, and J. E. Thomas
\href{https://doi.org/10.1103/PhysRevLett.113.020406}{Phys. Rev. Lett. {\bf 113}, 020406 (2014).}



\end{thebibliography}
\end{document}